\documentclass[a4paper,10pt]{article}
\usepackage{graphicx}
\usepackage{amsfonts}
\usepackage{amsthm}
\usepackage{nicefrac}
\usepackage{dcolumn}
\usepackage{multirow}
\usepackage{mathrsfs}
\usepackage{bbm}
\usepackage{hyperref}
\usepackage{comment}
\usepackage{float}
\usepackage{caption}
\usepackage{subcaption}
\usepackage{xcolor}
\usepackage{soul}
\usepackage{amsmath}
\usepackage{amssymb}
\usepackage{dcolumn}
\usepackage{bm}
\usepackage{cancel}
\usepackage[normalem]{ulem}

\renewcommand{\strut}{\vrule width 0pt height 12pt depth 8pt}
\newcommand{\struth}{\vrule width 0pt height 16 pt depth 16pt}
\newcommand{\strutl}{\vrule width 0pt height 0pt depth 5pt}
\renewcommand{\arraystretch}{2.25}
\usepackage{ifpdf} 
 
\begin{document}

\huge

\begin{center}
Distribution of angular momenta $M_L$ and $M_S$ in non-relativistic configurations: 
statistical analysis using cumulants and Gram-Charlier series
\end{center}

\vspace{0.5cm}

\large

\begin{center}
Jean-Christophe Pain$^{a,b,}$\footnote{jean-christophe.pain@cea.fr} and Michel Poirier$^{c}$
\end{center}

\normalsize

\begin{center}
\it $^a$CEA, DAM, DIF, F-91297 Arpajon, France\\
\it $^b$Universit\'e Paris-Saclay, CEA, Laboratoire Mati\`ere en Conditions Extr\^emes,\\
\it F-91680 Bruy\`eres-le-Ch\^atel, France\\
\it $^c$Universit\'e Paris-Saclay, CEA, LIDYL, F-91191 Gif-sur-Yvette, France
\end{center}

\vspace{0.5cm}

\begin{abstract}
The distributions $P(M_L,M_S)$ of the total magnetic quantum numbers $M_L$ and $M_S$ for $N$
electrons of angular momentum $\ell$, as well as the enumeration of $LS$ spectroscopic terms 
and spectral lines, are crucial for the calculation of atomic structure and spectra, in 
particular for the modeling of emission or absorption properties of hot plasmas. 
However, no explicit formula for $P(M_L,M_S)$ is known yet.
In the present work, we show that the generating function for the cumulants, which 
characterize the distribution, obeys a recurrence relation, similar to the Newton-Girard 
identities relating elementary symmetric polynomials to power sums. This enables us to 
provide an explicit formula for the generating function.
We also analyze the possibility of representing the $P(M_L,M_S)$ distribution by 
a bi-variate Gram-Charlier series, which coefficients are obtained from the knowledge of the exact moments of $P(M_L,M_S)$. 
It is shown that a simple approximation is obtained 
by truncating this series to the first few terms, though it is not convergent. 
\end{abstract}

\vspace{0.5cm}

\section{Introduction}\label{sec1}

In the non-relativistic case, electrons are characterized by uncoupled moments $\ell,s$, 
where $s$ is the one-half spin and $\ell$ the orbital quantum number. The magnetic quantum numbers $M_L$ and $M_S$ are respectively the sums of 
the individual numbers $m_\ell,m_s$ for each electron. The enumeration of the spectroscopic $LS$ terms arising in a given non-relativistic
configuration made of $\ell^N$ subshells, $N$ being the number of electrons, was addressed by different methods, such as the so-called vector 
model \cite{COWAN1981}, recurrence relations \cite{BAUCHE1987} or group theory \cite{BREIT1926,CURL1960,KARAYIANIS1965,KATRIEL1989,XU2006}. The knowledge of $LS$ terms, of the distribution $Q(L,S)$ of angular momenta $L$ and $S$, as well as of the distribution $P(M_L,M_S)$ of their projections ($M_L$ and $M_S$), is a prerequisite for the determination 
of the lines between two configurations, which plays a major role in the 
study of emission or absorption spectral properties of hot plasmas \cite{BAUCHE1988}, encountered 
for instance in stellar physics \cite{CLAYTON1983}, inertial-confinement fusion \cite{ATZENI2004}, or laser-plasma experiments \cite{KURZWEIL2022}. The latter applications imply taking into account complex ions, i.e., multi-electron configurations with several 
open subshells \cite{KRIEF2021}. The properties (regularities, trends) of $Q(L,S)$ and $P(M_L,M_S)$, are also worth investigating \cite{JUDD1968}, from a fundamental point of view but also in order to develop approximate models \cite{BAUCHE1991}. The statistics of electric-dipole (E1) lines 
was studied by Moszkowski \cite{MOSZKOWSKI1960}, Bancewicz \cite{BANCEWICZ1984}, Bauche and 
Bauche-Arnoult \cite{BAUCHE1987,BAUCHE1990}, Kucas \cite{KUCAS1995II,KUCAS1995III}, 
Gilleron and Pain \cite{GILLERON2009} and by us in a previous article \cite{POIRIER2021b}. 
Such a quantity is important for opacity codes, for instance, in order to decide whether a 
transition array can be described statistically or requires a detailed-line accounting calculation, 
relying on the diagonalization of the Hamiltonian \cite{PORCHEROT11}. In the same spirit, the statistics 
of electric quadrupole (E2) lines was also investigated \cite{PAIN2012}. We recently published 
explicit and recurrence formulas for magnetic quantum number $M_J$, projection the of the total angular momentum $\vec{J}=\vec{L}+\vec{S}$ on the $z$ axis \cite{POIRIER2021a}, 
together with a statistical analysis through the computation of cumulants. 
Recurrence relations obtained within this formalism lead to analytical expressions for 
the number of states of a given value of total magnetic number $M$ and number of fermions 
$N$. Explicit formulas have been given for this distribution for the values of $N$ up to 4 
\cite{POIRIER2021c} and then obtained up to $N=6$ together with a general algorithm for any number of fermions \cite{POIRIER2024}.
This formalism also allows one to obtain closed-form expressions for the
distribution of the total angular momentum $J$ \cite{POIRIER2021b}. Worth noticing is the expression of the 
total number of levels given by a piece-wise polynomial, derived in the above quoted papers \cite{POIRIER2021c} and \cite{POIRIER2024} 
for $N=3,4,5$ and 6 in the so-called ``relativistic 
configuration'' framework, i.e., when the electrons are characterized by their total angular 
momentum $j$. A particular case of fluctuation, 
the odd-even staggering (i.e., the fact that, in an electronic configuration, the number of odd values 
of $J$ can differ from the number of even values of $J$), was also studied \cite{BAUCHE1997,POIRIER2021b}.

The object of this work is to show that similar considerations apply to the distribution of the 
magnetic quantum numbers $M_L$ and $M_S$. Up to our knowledge, there exists no compact analytical expression for the 
quantum magnetic number distribution, which motivates the present investigations.

The paper is organized as follows. In section \ref{sec2}, the recursive determination of $P(M_L,M_S)$ is outlined. The latter method is a generalization of the one described in Ref.~\cite{GILLERON2009} (which was limited to $(J,M_J)$). In section \ref{sec3}, we introduce the two-variable generating functions for the cumulants of the ($M_L$, $M_S$) joint distribution. A recurrence relation for that function is derived. Based on an analogy with the Newton-Girard identities for elementary symmetric polynomials, an explicit formula is provided for the cumulant generating function. Finally, expressions of the first moments for small values of $\ell$ and of the number of fermions $N$ are provided in section \ref{sec4}, and the corresponding expressions
for the cumulants are given in section \ref{sec5}. In section \ref{sec6}, the exact
expressions for the magnetic quantum number distribution are
compared to simple analytical formulas based on the Gram-Charlier expansion.

It is worth noting that, with minor changes, the present work also concerns 
nuclear physics, where nucleons in a given shell are characterized by a total 
angular momentum $j$ and isospin $1/2$ \cite{ZHAO2003,ZHANG2008,Pain2019}.

\section{Numerical determination of $(L,S)$ statistics using a recursion relation for $P(M_L,M_S)$}\label{sec2}

The number $Q(L,S)$ of $LS$ terms of a configuration $\ell^N$ can be obtained from the relation 
\begin{align}
Q(L,S)&=\sum_{M_L=L}^{L+1}\sum_{M_S=S}^{S+1}(-1)^{L-M_L+S-M_S}P\left(M_L,M_S\right)\nonumber\\
&=P(L,S)-P(L+1,S)-P(L,S+1)+P(L+1,S+1),
\end{align}
where $P(M_L,M_S)$ is the number of states with total orbital magnetic 
number $M_L$ and spin magnetic number $M_S$. 
For a $N$-electron configuration $\ell_1^{N_1}\ell_2^{N_2}
\ell_3^{N_3}\cdots\ell_w^{N_w}$, $P(M_L,M_S)$ is determined through the 
standard convolution formula 
\begin{equation}\label{eq:convolM}
P\left(M_L,M_S\right)=\left(P_1\otimes P_2\otimes\cdots\otimes 
P_w\right)\left(M_L,M_S\right),
\end{equation}
$P_i$ being the $(M_L,M_s)$ distribution for the $i$-th subshell 
$\ell_i^{N_i}$. The distributions are convoluted two at a time, namely 
\begin{equation}
\left(P_i\otimes P_j\right)\left(M_L,M_S\right)=\sum_{M_L'}\sum_{M_S'}
 P_i\left(M_L',M_S'\right) P_j\left(M_L-M_L',M_S-M_S'\right).
\end{equation}

The individual-subshell distributions $P_i(M_L,M_S)$ can be obtained by 
an efficient algorithm proposed in Ref.~\cite{GILLERON2009}, adapted to 
the present situation, where $(M_L,M_S)$ are considered separately, 
instead of their sum $M_L+M_S$. The idea is to consider the $g=4\ell+2$ 
one-electron states (or micro-states) of $\ell^N$. Each of them is characterized by the projection $m_i=m_{\ell,i}+m_{s,i}$, with the spin projection $m_{s,i}=(-1)^i/2$ and the orbital-angular-momentum projection $m_{\ell,i}=\left[2i-4\ell-3-(-1)^i\right]/4$.
One has therefore $m_i=\left[2i-4\ell-3+(-1)^i\right]/4$. The index $i$ varies from 1 to $4\ell+2$. $P(M_L,M_S)$ is the number of $N$-electron states such as $m_{\ell,1}+\cdots m_{\ell,N}=M_L$ and $m_{s,1}+\cdots m_{s,N}=M_S$. If the last state (corresponding, in the previous ordering, to $i=4\ell+2$) is occupied by one electron (having therefore projections $m_{\ell,4\ell+2}=\ell$ and $m_{s,4\ell+2}=1/2$), then the $N-1$ remaining electrons must be distributed in the $4\ell+1$ remaining one-electron states, their total projections being then $M_L-m_{\ell,4\ell+2}=M_L-\ell$ and $M_S-m_{s,4\ell+2}=M_S-1/2$. This reads
\begin{equation}\label{GPMLMS}
P_{N,4\ell+2}(M_L,M_S)=P_{N-1,4\ell+1}(M_L-\ell,M_S-1/2)+P_{N,4\ell+1}(M_L,M_S).
\end{equation}
Such a reasoning applies for any subset of $k$ one-electron states, and one obtains the recurrence relation
\begin{equation}\label{recJCPMP}
P_{n,k}(M_L,M_S)=P_{n-1,k-1}(M_L-m_{\ell,k},M_S-m_{s,k})+P_{n,k-1}(M_L,M_S),
\end{equation}
where $P_{n,k}(M)$ represents the number of states with $n$ electrons
populating $k$ one-electron states, and giving projections $M_L$ and $M_S$. The recurrence is initialized with $P_{0,k}\left(M_L,M_S\right)=
\delta_{M_L,0}\delta_{M_S,0}
$ where $\delta_{i,j}$ is the Kronecker symbol
and $k$ is varied from 1 to $g=4\ell+2$ (total number of one-electron states), $n$ from 1 to $N$, $M_S$ from $-M_{S,\text{max}}$ to $M_{S,\text{max}}$ and $M_L$ from $-M_{L,\text{max}}$ to $M_{L,\text{max}}$. 

We have
\begin{equation}\label{smax}
M_{S,\text{max}}=
\begin{cases}N/2 & \text{ if } N\leq 2\ell+1,\\
(4\ell+2-N)/2 & \text{ if } N\geq 2\ell+1,\end{cases}
\end{equation}
and
\begin{equation}
M_{L,\text{max}}=\left[(-1)^N-1+2N(4\ell+2-N)\right]/8.
\end{equation}

\begin{table}[!ht]
\centering
\renewcommand{\arraystretch}{1.25}
\begin{tabular}{@{\quad}c@{\quad}*{13}{c}}\hline\hline
$L$ & 0 & 1 & 2 & 3 & 4 & 5 & 6 & 7 & 8 & 9 & 10 & 11 & 12 \\\hline
$Q(L,1/2)$ & 8 & 26 & 37 & 46 & 46 & 44 & 36 & 28 & 19 & 12 & 6 & 3 & 1 \\
$Q(L,3/2)$ & 7 & 18 & 28 & 32 & 33 & 29 & 24 & 17 & 11 & 6 & 3 & 1 & 0 \\
$Q(L,5/2)$ & 1 & 5 & 5 & 8 & 6 & 6 & 4 & 3 & 1 & 1 & 0 & 0 & 0 \\
\hline\hline
\end{tabular}
\caption{Value of the number of $LS$ spectroscopic terms $Q(L,S)$ for the configuration $\text{d}^2\text{f}^3$ and three values of $S$: 1/2, 3/2, and 5/2. The total degeneracy of $\text{d}^2\text{f}^3$ is $\binom{10}{2}\binom{14}{3}$ = 16380, the maximum orbital angular momentum is $L_{\text{max}}$=12 and the maximum spin $S_{\text{max}}$=5/2. The correctness of the results can be checked using $\sum_{M_L,M_S} P(M_L,M_S)$ = 16380 and $\sum_{L,S}(2L+1)(2S+1)Q(L,S)$=16380. The total number of $LS$ spectroscopic terms 
 is $\sum_{L,S}Q(L,S)$=561.}\label{tab1}
\end{table}

Table \ref{tab1} provides values of $Q(L,S)$ for the $\text{d}^2\text{f}^3$ configuration and $S$=1/2, 3/2 and 5/2. For fixed $S$, $Q(L,S)$ has a truncated bell shape.

We may estimate the number of operations needed to obtain the whole set of 
$P(M_L,M_S)$ values in a $\ell^N$ subshell. The brute-force technique 
consists in evaluating all the
\begin{equation}\label{eq:nop_bf}
N_\text{bf}=\binom{4\ell+2}{N}
\end{equation}
$N$-tuple elements and compute the sums $\sum_{i=1}^N m_{\ell,i}$ and $\sum_{i=1}^N m_{s,i}$ for each of them. The much better alternative provided by the recurrence (\ref{recJCPMP}) amounts to perform roughly
\begin{equation}
N_{\mathrm{Eq.} (\ref{recJCPMP})}=N(4\ell+2)(2M_{L,\mathrm{max}}+1)(2M_{S,\mathrm{max}}+1)
\end{equation}
operations: (number of one-electron states) $\times$ (number of electrons) $\times$ (number of values of $M_L$) $\times$ (number of values of $M_S$). Thus, if $N\leq 2\ell+1$:
\begin{equation}
N_{\mathrm{Eq.} (\ref{recJCPMP})}=N(N+1)(4\ell+2)\left[\frac{(-1)^N+3+2N(4\ell+2-N)}{4}\right]
\end{equation}
and if $N\geq 2\ell+1$
\begin{equation}
N_{\mathrm{Eq.} (\ref{recJCPMP})}=N(4\ell+3-N)(4\ell+2)\left[\frac{(-1)^N+3+2N(4\ell+2-N)}{4}\right].
\end{equation}
For a half-filled subshell, one has
\begin{equation}
N_{\mathrm{Eq.} (\ref{recJCPMP})}=4(\ell+1)(2\ell+1)^2\left[2\ell(\ell+1)+1\right].
\end{equation}

\begin{table}[htbp]
\renewcommand{\arraystretch}{1.5}
\centering
\begin{tabular}{@{\quad}c@{\quad}*{9}{c}}\hline\hline
$\ell$           & 0 & 1 & 2 & 3 & 4 & 5 & 6 & 7 & 8\\\hline
$N_{\mathrm{Eq.} (\ref{recJCPMP})}$ & 4 & 360 & 3900 & 19600 & 66420 & 177144 & 402220 & 813600 & 1508580 \\
 $N_{\text{bf}}$ & 2 & 20 & 252 & 3432 & 48620 & 705432 & 10400600 & 155117520 & 2333606220 \\\hline\hline
\end{tabular}
\caption{Number of operations needed to obtain the $P(M_L,M_S)$ distribution 
for the $\ell^N$ configuration with $N=2\ell+1$, using a brute-force technique 
or recurrence relation (\ref{recJCPMP}).}\label{tab2}
\end{table}

Table \ref{tab2} shows the number of operations needed to obtain the $P(M_L,M_S)$ distribution 
for the $\ell^N$ configuration with $N=2\ell+1$, using a brute-force technique 
or recurrence relation (\ref{recJCPMP}). We can see that for $\ell>4$, the recurrence relation becomes much more efficient than the direct calculation.

\section{The two-variable generating function for the cumulants of the ($M_L$, $M_S$) joint 
distribution}\label{sec3}

\subsection{Definition}\label{subsec31}

As stated in statistical treatises \cite{STUART1994}, the whole information 
about the distribution $P(M_L,M_S)$ of magnetic quantum number is contained in 
the exponential of the cumulant generating function defined as 
\begin{equation}
 \exp(K(u,v)) = \left<\exp(uM_L+vM_S)\right>
 = \sum_{M_L,M_S} P(M_L,M_S)\,e^{uM_L+vM_S} \left/ \sum_{M_L,M_S} P(M_L,M_S). \right.\label{eq:defeKt}
\end{equation}
From the Pauli exclusion principle this normalization factor is given by the binomial coefficient
\begin{equation}
\sum_{M_L,M_S} P(M_L,M_S)=\binom{4\ell+2}{N}.\label{eq:norm}
\end{equation}
The numerator of the cumulant generating function of the distribution $P(M_L,M_S)$ reads
\begin{equation}\label{eq:defs}
s(N,\ell,u,v)=\sum_{M_L,M_S}P(M_L,M_S)\,e^{uM_L}e^{vM_S}
=\binom{4\ell+2}{N}\exp(K(u,v)),
\end{equation}
and will be, for simplicity, referred to as the ``generating function'' in the following. 

\subsection{Cumulant generating function for the first values of $N$}\label{subsec32}

The $N=1$ value is given by
\begin{equation}
s(1,\ell,u,v)=\sum_{m_{\ell}=-\ell}^{\ell}\sum_{m_s=-1/2}^{1/2}e^{m_{\ell}u}e^{m_{s}v}=2\frac{\sinh\left[(2\ell+1)u/2\right]}{\sinh(u/2)}\cosh(v/2).\label{eq:vals1}
\end{equation}
The generating function in the case of two fermions reads
\begin{align}
s(2,\ell,u,v)&=\frac{1}{2}\left[\sum_{k=1}^ge^{m_{\ell,k}u}e^{m_{s,k}v}\sum_{i=1, i\ne k}^ge^{m_{\ell,i}u}e^{m_{s,i}v}\right]\nonumber\\
&=\frac{1}{2}\left[\sum_{k=1}^ge^{m_{\ell,k}u}e^{m_{s,k}v}\left(\sum_{i=1}^ge^{m_{\ell,i}u}e^{m_{s,i}v}-e^{m_{\ell,k}u}e^{m_{s,k}v}\right)\right]\nonumber\\
&=\frac{1}{2}\left[\left(\sum_{k=1}^ge^{m_{\ell,k}u}e^{m_{s,k}v}\right)^2-\sum_{k=1}^ge^{2m_{\ell,k}u}e^{2m_{s,k}v}\right]\nonumber\\
&=\frac{1}{2}\left[\left(\sum_{k=1}^ge^{m_{\ell,k}u+m_{s,k}v}\right)^2-\sum_{k=1}^ge^{2m_{\ell,k}u+2m_{s,k}v}\right].
\end{align}

Let us set
\begin{equation}\label{eq:Sp}
\mathscr{S}_p=\sum_{k=1}^ge^{p(m_{\ell,k}u+m_{s,k}v)}=2\frac{\sinh\left[(2\ell+1)pu/2\right]}{\sinh(pu/2)}\cosh(pv/2).
\end{equation}
By comparison with Eq. (\ref{eq:vals1}), one can check that $\mathscr{S}_p=s(1,\ell,pu,pv)$. 
We have also
\begin{equation}
s(2,\ell,u,v)=\frac{1}{2}\left(\mathscr{S}_1^2-\mathscr{S}_2\right),
\end{equation}
and a similar calculation for $N=3$ gives
\begin{equation}\label{Neq3}
s(3,\ell,u,v)=\frac{1}{6}\left(\mathscr{S}_1^3-3\mathscr{S}_1\mathscr{S}_2+2\mathscr{S}_3\right).
\end{equation}
In the case of four fermions, a similar direct calculation yields
\begin{equation}\label{s4}
s(4,\ell,u,v)=\frac{1}{24}\left(\mathscr{S}_1^4-6\mathscr{S}_1^2\mathscr{S}_2+3\mathscr{S}_2^2+8\mathscr{S}_1\mathscr{S}_3-6\mathscr{S}_4\right).
\end{equation}

\subsection{Recurrence over the number of fermions for the cumulant generating function}\label{subsec33}

The generating function can be expressed as a multiple summation over one-electron states
\begin{multline}
s(N,\ell,u,v)=\sum_{M_L,M_S}\sum_{p_1=0}^1\sum_{p_2=0}^1\cdots\sum_{p_g=0}^1 \\
 \delta\left(M_L-\sum_{k=1}^{g}p_km_{\ell,k}\right)\, \delta\left(M_S-\sum_{k=1}^{g}p_km_{s,k}\right)\,\delta\left(N-\sum_{k=1}^{g}p_k\right)e^{M_Lu}e^{M_Sv},
\end{multline}
where the sum over $M_L$ and $M_S$ may be eliminated
\begin{equation}
s(N,\ell,u,v)=\sum_{p_1=0}^1\sum_{p_2=0}^1\cdots\sum_{p_g=0}^1 
 \delta\left(N-\sum_{k=1}^g p_k\right) \exp\left(\sum_{k=1}^g p_km_{\ell,k}u\right) \exp\left(\sum_{k=1}^g p_km_{s,k}v\right).
\end{equation}
In the above equations, for the sake of readability 
$\delta(Q)$ stands for the Kronecker symbol $\delta_{Q,0}$.
Isolating in this multiple sum the contributions of the $p_g$ index and then the $p_1$ index, repeating exactly the procedure used in Eq. (2.14) of \cite{POIRIER2021a}, one gets the recurrence property on the generating function 
\begin{equation}
s(N,\ell,u,v)=s(N,\ell-1,u,v)+2\cosh\left(\ell u+v/2\right)s(N-1,\ell-1,u,v)+s(N-2,\ell-1,u,v).
\label{eq:recj}
\end{equation}
The recurrence (\ref{eq:recj}) is the same as for a single moment, \emph{i.e.} (5.5) of Ref. \cite{POIRIER2021a}. Thus, in the present case, the recurrence over the number of fermions only is, \emph{mutatis mutandis}, 
\begin{equation}
s(N,\ell,u,v) = \frac{1}{N}\sum\limits_{p = 1}^N {{{\left( { - 1} \right)}^{p + 1}}s(1,\ell,pu,pv)} \,{s(N-p,\ell,u,v)},\label{eq:recNfngen}
\end{equation}
with $s(0,\ell,u,v)=1$. The case of a general spin (non necessarily $1/2$), i.e., the generalization to a pair of arbitrary angular momenta, is discussed in appendix \ref{appA}.

\subsection{General expression of the generating function}\label{subsec34}

Let $x_1$, ..., $x_n$ be variables and denote for $k\ge1$ by $p_k(x_1, ..., x_n)$ the $k-$th power sum:
\begin{equation}
 p_k(x_1,\ldots ,x_n) = \sum_{i=1}^n x_i^k = x_1^k+\cdots +x_n^k,
\end{equation}
and for $k\ge 0$ denote by $e_k(x_1, ..., x_n)$ the elementary symmetric polynomial (that is, 
the sum of all distinct products of $k$ distinct variables). One has
\begin{align}
e_k(x_1,\cdots,x_n) &=\sum_{1\le i_1<i_2<\cdots<i_k\le n} 
x_{i_1}x_{i_2}\cdots x_{i_k},
\end{align}
and $e_0=1, e_k=0$ if $k > n$. The Newton-Girard identities can be stated as (see for instance Refs. \cite{MACDONALD1979,MEAD1992,KALMAN2000}):
\begin{equation}
ke_k(x_1,\ldots ,x_n) = \sum_{i=1}^k (-1)^{i-1} e_{k-i}(x_1,\ldots ,x_n)\,p_i(x_1,\ldots ,x_n),\label{eq:Newtonid}
\end{equation}
valid for all $n\ge 1$ and $n \ge k \ge 1$. It can be shown that
\begin{equation}
e_n = (-1)^n \sum_{\stackrel{m_1+2m_2+\cdots+nm_n=n}{m_1\ge0,\,\cdots, \,m_n\ge0}} \;\prod_{i=1}^n \frac{(-p_i)^{m_i}}{m_i!\,i^{m_i}}.
\end{equation}
Our recurrence relation over the number of fermions (\ref{eq:recNfngen}) is the same as Eq. (\ref{eq:Newtonid}), making the replacements $e_N\rightarrow s(N,\ell,u,v)$ and $p_i\rightarrow\mathscr{S}_i$. For the $N$-fermion case, we have therefore, {\it mutatis mutandis}, the general expression
\begin{equation}\label{eq:sNsum}
s(N,\ell,u,v)=(-1)^N\sum_{\vec{q}/\\q_1+2q_2+\cdots+Nq_N=N}\prod_{p=1}^N\frac{1}{q_p!}\left(-\frac{\mathscr{S}_p}{p}\right)^{q_p},
\end{equation}
where $\vec{q}$ stands for the $N$-fold set $(q_1,q_2,\dots q_N)$ 
and with
\begin{equation}\label{eq:exprSp}
\mathscr{S}_p=2\frac{\sinh\left[(2\ell+1)pu/2\right]}{\sinh(pu/2)}\cosh(pv/2).
\end{equation}
A more detailed proof is provided in appendix \ref{appB}. The numerical implementation of Eq. (\ref{eq:sNsum}) requires the numerical determination of partitions of an integer.

In addition, the cumulant generating function $s(N,\ell,u,v)$ itself can be expressed in terms of incomplete Bell polynomials $B_{n,k}$. One has
\begin{equation}
s(N,\ell,u,v)=\frac{1}{N!}\sum_{k=1}^NB_{N,k}(\mathscr{S}_1,-\mathscr{S}_2,\cdots, (-1)^{p-1}(p-1)!\mathscr{S}_p,\cdots,(-1)^{N-k}(N-k)!\mathscr{S}_{N-k+1})
\end{equation}
where
\begin{equation}\label{defbell}
B_{n,k}(x_1,x_2,\cdots,x_{n-k+1})=\sum \frac{n!}{j_1!j_2!\cdots j_{n-k+1}!}\left(\frac{x_1}{1!}\right)^{j_1}\left(\frac{x_2}{2!}\right)^{j_2}\cdots\left(\frac{x_{n-k+1}}{(n-k+1)!}\right)^{j_{n-k+1}},
\end{equation}
the sum running over all ensembles of integers $j_1, j_2, \cdots, j_{n-k+1}$ such that $j_1+j_2+\cdots+j_{n-k+1}=k$ and $j_1+2j_2+3j_3+\cdots+(n-k+1)j_{n-k+1}=n$.

\section{Determination of the first moments for 2 and 3 fermions}\label{sec4}

The expansion of $s(N,\ell,u,v)/\binom{4\ell+2}{N}$ at $(u=0,v=0)$ 
provides the moments $\mu_{m,n}$.
The moments $\mu_{m,n}$ are defined by the expansion 
\begin{equation}\label{eq:defM}
M(u,v)=e^{K(u,v)}=\sum_{m,n=0}^{\infty}\mu_{m,n}\frac{u^mv^n}{m!n!},
\end{equation}
where $M(u,v)$ is proportional to the generating function defined by 
Eq.~(\ref{eq:defs}).
The formulas of the moments can be obtained easily using a computer algebra system, such as Mathematica.

\subsection{Moments for $N=2$}\label{subsec41}

For $N=2$, one has
\begin{equation}
s(2,\ell,u,v)=\frac{\sinh\left[(2\ell+1)u/2\right]}{\sinh(u/2)\sinh u}\left\{(1+2\cosh v)\sinh(\ell u)+\sinh[(\ell+1)u]\right\}.
\end{equation}
The expressions of the moments as functions of $\ell$ for $N=2$ and $m=0,2$, are given in Table \ref{tab3c}. 

\renewcommand{\strutl}{\vrule width 0pt height 0pt depth 12pt}
\begin{table}[htbp]
\centering
\begin{tabular}{c@{\quad}*{2}{c}}\hline\hline
$m$ & $n=0$ & $n=2t$\\\hline
0   & 1 & $\dfrac{2\ell}{(4\ell+1)}$ \\
2   & $\dfrac{8\ell^2(\ell+1)}{3(4\ell+1)}$ & $\dfrac{2\ell(\ell+1)(2\ell-1)}{3(4\ell+1)}$ \strutl\\
\hline\hline
\end{tabular}
\caption{Expressions of the moments $\mu_{m,n}$ for $N=2$, $m=0,2$ and $n=2t$ with $t>0$. 
The moments $\mu_{n,2t}$ are independent of $t$ if $t>0$.}\label{tab3c}
\end{table}

For $m=4$, one has, $t$ being any positive integer,
\begin{subequations}\begin{align}
\mu_{4,0}(\ell^2)&=\frac{4\ell(\ell+1)(16\ell^3+12\ell^2-6\ell+3)}{15(4\ell+1)},\\
\mu_{4,2t}(\ell^2)&=\frac{2\ell(\ell+1)(2\ell-1)(8\ell^2+4\ell-7)}{15(4\ell+1)}.
\end{align}\end{subequations}

It is easy to show that $\mu_{2s,2t}$ is independent of $t>0$ in the two-electron case. 
Indeed, one has then $S=0$ or 1, so that $M_S=0,\pm1$ and $M_S^{2t}=M_S^2$ does not 
depend on $t$, the same property being valid for $\mu_{4,2t}$. 
For $n=6$, $t$ being again any positive integer, one gets
\begin{subequations}\begin{align}
\mu_{6,0}(\ell^2)&=\frac{4\ell(\ell+1)(48\ell^5+72\ell^4-20\ell^3-6\ell^2+40\ell-15)}{21(4\ell+1)},\\
\mu_{6,2t}(\ell^2)&=\frac{2\ell(\ell+1)(2\ell-1)(24\ell^4+24\ell^3
-46\ell^2-26\ell+31)}{21(4\ell+1)},
\end{align}\end{subequations}
and $\mu_{m,n}(\ell^2)=0$ if $n$ is odd. 

\subsection{Moments for $N=3$}\label{subsec42}

The general expressions of moments for $N=3$ are more cumbersome. One has
\begin{multline}
s(3,\ell,u,v)=\frac{\sinh(\ell u)\sinh[(2\ell+1)u/2]\sinh\big((\ell+1)u\big)}{\sinh(u/2)\sinh (u)\sinh (3u/2)}~U_1(\cosh(u/2))U_1(\cosh(v/2))\\
 +\frac{\sinh\big((2\ell-1)u/2\big)\sinh(\ell u)\sinh[(2\ell+1)u/2]}{\sinh(u/2)\sinh(u)\sinh(3u/2)}~U_3(\cosh(v/2)),
\end{multline}
where $U_n$ is the Chebyshev polynomial of the second kind of order $n$. 
Explicitly, one has $U_1(X)=2X$, $U_3(X)=8X^3-4X$.
The expressions of the moments as functions of $\ell$ for $N=3$, $m=0,2$ and $n=0,2,4$ are given in Table \ref{tab3bm}. 

\begin{table}[htbp]
\centering
\begin{tabular}{c@{\quad}*{3}{c}}\hline\hline\strut
$m$ & $n=0$ & $n=2$ & $n=4$\\\hline\struth
0   & 1 & $\dfrac{3(4\ell-1)}{4(4\ell+1)}$ & $\dfrac{3(28\ell-13)}{16(4\ell+1)}$\\
2   & $\dfrac{\ell(\ell+1)(4\ell-1)}{(4\ell+1)}$ & $\dfrac{(\ell+1)(12\ell^2-13\ell+4)}{4(4\ell+1)}$ & $\dfrac{(\ell+1)(84\ell^2-121\ell+40)}{16(4\ell+1)}$\strutl\\\hline\hline
\end{tabular}
\caption{Expressions of the moments $\mu_{m,n}$ for $N=3$, $m=0,2$ and $n$ varying from 0 to 4. The moments are zero if $m$ or $n$ is odd.}\label{tab3bm}
\end{table}

For $m=4$, one has
\begin{subequations}\begin{align}
\mu_{4,0}(\ell^3)&=\frac{(\ell+1)(52\ell^4+13\ell^3-35\ell^2+21\ell-6)}{5(4\ell+1)},\\
\mu_{4,2}(\ell^3)&=\frac{(\ell+1)(156\ell^4-143\ell^3-135\ell^2+249\ell-82)}{20(4\ell+1)},\\
\mu_{4,4}(\ell^3)&=\frac{(\ell+1)(1092\ell^4-1547\ell^3-1035\ell^2+2301\ell-766)}{80(4\ell+1)},
\end{align}\end{subequations}
and $\mu_{m,n}(\ell^3)=0$ if $n$ is odd.

\section{Expressions of the first cumulants in the two- and 
three-fermion cases}\label{sec5}

The cumulants $\kappa_{m,n}$ are obtained by expanding  $\log(s(N,\ell,u,v)/\binom{4\ell+2}{N})$ at $(u=0,v=0)$.

\subsection{Cumulants for $N=2$}

The expressions of the cumulants as functions of $\ell$ for $N=2$, are given in Table \ref{tab3} for $m=0,2$ and $n=0, 2$ and 4.

\renewcommand{\strutl}{\vrule width 0pt height 0pt depth 12pt}
\begin{table}[htbp]
\centering
\begin{tabular}{c@{\quad}*{3}{c}}\hline\hline
$m$ & $n=0$ & $n=2$ & $n=4$\\\hline
0   & 0 & $\dfrac{2\ell}{4\ell+1}$ & $-\dfrac{2\ell(2\ell-1)}{(4\ell+1)^2}$\\
2   & $\dfrac{8\ell^2(\ell+1)}{3(4\ell+1)}$ & $-\dfrac{2\ell(\ell+1)(2\ell+1)}{3(4\ell+1)^2}$ & $\dfrac{2\ell(\ell+1)(2\ell+1)(8\ell-1)}{3(4\ell+1)^3}$\strutl\\
\hline\hline
\end{tabular}
\caption{Expressions of the cumulants $\kappa_{m,n}$ for $N=2$, $m=0,2$ and $n$ varying from 0 to 4.}\label{tab3}
\end{table}

For $m=4$, one has
\begin{subequations}\begin{align}
    \kappa_{4,0}(\ell^2)&=-\frac{4\ell(\ell+1)(2\ell-1)(8\ell^3+12\ell^2+12\ell+3)}{15(4\ell+1)^2},\\
    \kappa_{4,2}(\ell^2)&=\frac{2\ell(\ell+1)(2\ell+1)(16\ell^3+12\ell+7)}{15(4\ell+1)^3},\\
    \kappa_{4,4}(\ell^2)&=-\frac{2\ell(\ell+1)(2\ell+1)(128\ell^4+104\ell^3+276\ell^2+104\ell-7)}{15(4\ell+1)^4},
\end{align}\end{subequations}
and $\kappa_{m,n}(\ell^2)=0$ if $n$ is odd.

\subsection{Cumulants for $N=3$}

The expressions of the cumulants as functions of $\ell$ for $N=3$, are given in Table \ref{tab3b}. 

\begin{table}[htbp]
\centering
\begin{tabular}{c@{\quad}*{4}{c}}\hline\hline\strut
$m$ & $n=0$ & $n=2$ & $n=4$\\\hline\struth
0   & 0 & $\dfrac{3(4\ell-1)}{4(4\ell+1)}$ & $-\dfrac{3(16\ell^2-24\ell+11)}{8(4\ell+1)^2}$\\
2   & $\dfrac{\ell(\ell+1)(4\ell-1)}{(4\ell+1)}$ & $-\dfrac{(\ell+1)(4\ell^2-1)}{(4\ell+1)^2}$ & $\dfrac{(\ell+1)(4\ell^2-1)(8\ell-7)}{(4\ell+1)^3}$\strutl\\\hline\hline
\end{tabular}
\caption{Expressions of the cumulants $\kappa_{m,n}$ for $N=3$, $m=0,2$ and $n$ equal to 0, 2 and 4. The cumulants are zero if $m$ or $n$ is odd.}\label{tab3b}
\end{table}

For $m=4$, one has
\begin{subequations}\begin{align}
    \kappa_{4,0}(\ell^3)&=-\frac{(\ell+1)(32\ell^5+16\ell^4+22\ell^3-34\ell^2+3\ell+6)}{5(4\ell+1)^2},\\
    \kappa_{4,2}(\ell^3)&=\frac{(\ell+1)(4\ell^2-1)(16\ell^3+8\ell^2+56\ell+25)}{5(4\ell+1)^3},\\
    \kappa_{4,4}(\ell^3)&=-\frac{(\ell+1)(4\ell^2-1)(128\ell^4+312\ell^3+752\ell^2-282\ell-265)}{5(4\ell+1)^4},
\end{align}\end{subequations}
and $\kappa_{m,n}(\ell^3)=0$ if $n$ is odd. 

\subsection{Relation between cumulants and moments for the first orders}

The cumulants $\kappa_{m,n}$ are obtained from the expansion 
\begin{equation}
K(u,v)=\sum_{m,n}\kappa_{m,n}\frac{u^mv^n}{m!n!},
\end{equation}
with $M(u,v)=\exp(K(u,v))$ (see Eq.~(\ref{eq:defM})).
For the $M_L,M_S$ distribution the function $K$ contains only even-order terms.

The relations between moments and cumulants in the bi-variate case can be 
obtained by the formal method devised by Kendall \cite{Kendall1940} and 
used by Cook \cite{Cook1951} who provided tables. For $m+n\le6$ one gets
\begin{subequations}\begin{align}
\kappa_{2,0}&=\mu_{2,0},
\\\kappa_{4,0}&=\mu_{4,0}-3\mu_{2,0}^2,
\\\kappa_{2,2}&=\mu_{2,2}-\mu_{2,0}\mu_{0,2}\\
\kappa_{6,0}&=\mu_{6,0}-15\mu_{2,0}\mu_{4,0}+30\mu_{2,0}^3,
\\\kappa_{4,2}&=\mu_{4,2}-\mu_{4,0}\mu_{0,2}-6\mu_{2,0}\mu_{2,2}
+6\mu_{2,0}^2\mu_{0,2}.
\end{align}\end{subequations}
These relations are completed using the symmetry $m\leftrightarrow n$, 
e.g., $\kappa_{0,4}=\mu_{0,4}-3\mu_{0,2}^2$, etc.

One can also express the moments $\mu_{m,n}$ as a function of cumulants
\begin{subequations}\begin{align}
\mu_{4,0}&=\kappa_{4,0}+3\kappa_{2,0}^2,\\
\mu_{2,2}&=\kappa_{2,2}+\kappa_{2,0}\kappa_{0,2}\\
\mu_{6,0}&=\kappa_{6,0}+15\kappa_{2,0}\kappa_{4,0}+15\kappa_{2,0}^3,\\
\mu_{4,2}&=\kappa_{4,2}+\kappa_{4,0}\kappa_{0,2}+6\kappa_{2,0}\kappa_{2,2}
+3\kappa_{2,0}^2\kappa_{0,2}.
\end{align}\end{subequations}

\section{Statistical modeling of the $P(M_L,M_S)$ distribution using Gram-Charlier expansion series}\label{sec6}

\subsection{Bi-variate Gram-Charlier series}

The Gram-Charlier expansion was derived in an attempt to express
non-Gaussian distributions as infinite series using the 
moments as input terms \cite{KENDALL1977}. The one-dimensional Gram-Charlier series has been widely used 
in different fields of physics. For instance, it was successfully applied to the statistical modeling of transition arrays of absorption or emission lines in hot-plasma complex spectra, or of the distribution of angular momentum $M_J$ in atomic configurations. 
The two-variable (or bi-variate) Gram-Charlier series are much less frequent in the 
literature. However, Kamp\'e de F\'eriet \cite{KAMPE1966} has provided 
formulas relevant for this case. If the two variables are uncorrelated 
(indeed we have here $\langle M_L M_S\rangle=0$), these expressions become 
much simpler, and one has 
\begin{subequations}
\begin{gather}
\mathscr{G}_T(u,v)=\frac{e^{-u^2/2\sigma^2-v^2/2\tau^2}}{2\pi\sigma\tau}
\left[1+\sum_{n=2}^T\sum_{j=0}^n c_{2n-2j,2j}\,He_{2n-2j}
\left(\frac{u}{\sigma}\right)He_{2j}\left(\frac{v}{\tau}\right)\right]
\label{eq:GCuv}\\
\text{where }u=M_L, v=M_S.
\end{gather}
\end{subequations}
In the above equations, $T$ is an arbitrary integer --- hereafter called ``half truncation 
order'' --- and $He_n(X)$ is the Hermite polynomial of order $n$ 
\cite{ABRAMOWITZ1972}. The approximation $\mathscr{P}_T(u,v)$ for 
$P(M_L,M_S)$ is obtained by multiplying the function $\mathscr{G}_T(u,v)$ 
by the configuration degeneracy 
\begin{equation}
\mathscr{P}_T(u,v)=\mathscr{N}\mathscr{G}_T(u,v)\quad
\text{ with }\mathscr{N}=\prod_{i=1}^w \binom{4\ell_i+2}{p_i}\label{eq:GCuvN}
\end{equation}
assuming that the configuration is made of $w$ subshells of orbital momentum 
$\ell_i$ and population $p_i$. The two variances entering Eq.~(\ref{eq:GCuv}) are given by $\sigma^2=\kappa_{2,0}$ and $\tau^2=\kappa_{0,2}$. 
In the present case, all the odd moments are zero
\begin{equation}
\left<u^{2p+1}\right>=0,\quad\left<v^{2q+1}\right>=0,
\end{equation}
as well as 
$\left<u^{2p+1}v^n\right>=0$, $\left<u^mv^{2q+1}\right>=0$.

The global error on all $M_L,M_S$ values may be characterized by 
the averages 
\begin{gather}\label{eq:abserr}
\Delta_\text{abs}(T)=\left(\left.\sum_{M_L,M_S}(\mathscr{P}_T(M_L,M_S)-
P(M_L,M_S))^2\right/ (2L_\text{max}+1)(2S_\text{max}+1)\right)^{1/2},\\
\label{eq:relerr}
\Delta_\text{rel}(T)=\left(\left.\sum_{M_L,M_S}{}^\prime
\:\Big(\mathscr{P}_T(M_L,M_S)/
P(M_L,M_S)\Big)^2\right/ N_\text{pos}\right)^{1/2},
\end{gather}
where the prime means that the sum is restricted to elements $(M_L,M_S)$ 
for which $P(M_L,M_S)>0$ and $N_\text{pos}$ is the number of such elements.
In the following discussion, we have chosen to compute both 
$\Delta_\text{abs}(T)$ and $\Delta_\text{rel}(T)$ since they convey 
different information. The absolute error is more sensitive to the 
differences where $P(M_L,M_S)$ is maximum, namely for $M_L,M_S$ close 
to 0, while the relative error favors regions where $P(M_L,M_S)$ is small, 
namely $|M_L|$ and/or $|M_S|$ close to their maximum value.  
In the following, for simplicity reasons, the first data point in 
plots for such errors is at $T=1$ which is identical to $T=0$. 
As seen in Eq.(\ref{eq:GCuv}), if $T<2$ the sum over $n$ is absent and 
the Gram-Charlier approximation simplifies into the Gaussian expression. 

\subsection{Moments expressed as functions of Gram-Charlier coefficients}

With the above expression (\ref{eq:GCuv}), one may express the various moments 
$\langle u^{2s}v^{2t}\rangle=\mu_{2s,2t}$ as a function of the Gram-Charlier coefficients $c_{2i,2j}$
\begin{subequations}\begin{align}
\left<u^2\right>&=\sigma^2,\\
\left<u^2v^2\right>&=\sigma^2\tau^2(1+4c_{2,2}),\\
\left<u^4\right>&=3\sigma^4(1+8c_{4,0}),\\
\left<u^4v^2\right>&=3\sigma^4\tau^2(1+8c_{2,2}+8c_{4,0}+16c_{4,2}),\\
\left<u^6\right>&=15\sigma^6(1+24c_{4,0}+48c_{6,0}),\\
\left<u^4v^4\right>&=9\sigma^4\tau^4(1+8c_{4,0}+16c_{2,2}+8c_{0,4}
+32c_{4,2}+32c_{2,4}+64c_{4,4}),\\
\left<u^6v^2\right>&=15\sigma^6\tau^2(1+12c_{2,2}+24c_{4,0}+48c_{4,2}+48c_{6,0}
+96c_{6,2}),\\
\left<u^8\right>&=105\sigma^8(1+48c_{4,0}+192c_{6,0}+384c_{8,0}),
\end{align}\end{subequations}
together with expressions obtained by changing $u\leftrightarrow v$. Such formulas assume that $T$ is sufficiently large, for instance if $T=6$, the terms $c_{6,2}$, 
$c_{4,4}$ and $c_{8,0}$ are absent from the three above formulas. 

It is possible to provide a general form of the moments from properties of the Hermite polynomials. 
We use the expansion (\ref{eq:GCuv}) and calculate the integral
$\left<u^pv^q\right>=\int_{-\infty}^{+\infty}du\,dv\; u^p v^q \mathscr{P}_T(u,v).$ 
Noting that
\begin{equation}
He_n(x)=n!\sum_{m=0}^{\lfloor\frac{n}{2}\rfloor}\frac{(-1)^m}{m!(n-2m)!}\frac{x^{n-2m}}{2^m},\label{eq:devHe}
\end{equation}
where $\lfloor x\rfloor$ is the integer part of $x$, the calculation of the average of $u^pv^q$ boils down to the evaluation of simple integrals of the kind
\begin{equation}\label{eq:defkhi}
\chi_p(\sigma)=\int_{-\infty}^{+\infty}du\; u^p e^{-u^2/2\sigma^2}.
\end{equation}
If $p$ is odd, one has $\chi_{p}(\sigma)=0$. 
For $p=2j$ one gets after basic algebra\begin{equation}
\chi_{2j}(\sigma)=\sqrt{2\pi}\sigma^{2j+1}\frac{(2j)!}{2^j j!}
\quad\text{if $p=2j$ is even}.
\end{equation}

From the distribution $\mathscr{G}_T$ (\ref{eq:GCuv}), the expansion 
(\ref{eq:devHe}) and the above definition of $\chi$, we get the average
\begin{align}\langle u^pv^q\rangle
&=\frac{1}{2\pi\sigma\tau}\left[\chi_p(\sigma)\chi_q(\tau)
+\sum_{n=2}^T\sum_{j=0}^n c_{2n-2j,2j}\,
\mathscr{V}_{p,2n-2j}(\sigma)\mathscr{V}_{q,2j}(\tau)\right]
\end{align}
with\begin{equation}
\mathscr{V}_{m,n}(\sigma)=\int_{-\infty}^{+\infty}\!\!du\;
u^m e^{-u^2/2\sigma^2}He_{n}(u/\sigma).
\end{equation}
Obviously $\mathscr{V}_{m,n}(\sigma)=0$ if $m+n$ is odd. 
After basic algebraic manipulations, one gets
$\mathscr{V}_{2s,2t}(\sigma)=0$ if $s<t$, and
\begin{equation}
\mathscr{V}_{2s,2t}(\sigma)=\frac{(2\pi)^{1/2}(2s)!}{ 2^{s-t}(s-t)!}\sigma^{2s+1} 
 \quad\text{if }s\ge t.
\end{equation}

\subsection{Gram-Charlier coefficients expressed in terms of averages}

Using the orthogonality relation
\begin{equation}
\int_{-\infty}^{\infty}{He}_{m}(x){He}_{n}(x)\,e^{-x^{2}/2}\,dx={\sqrt{2\pi}}\,n!\,\delta_{n,m},
\end{equation}
we get, after multiplying Eq. (\ref{eq:GCuv}) by 
$He_{2n-2j}(u/\sigma)He_{2j}(v/\tau)$ and integrating over $u,v$, 
\begin{equation}\label{cst}
c_{2n-2j,2j}=\sum_{k=0}^{n-j}\sum_{r=0}^{j}\frac{(-1)^{k+r}\sigma^{2k-2n+2j}\,\tau^{2r-2j}}{2^{k+r}k!r!(2n-2j-2k)!(2j-2r)!}\langle u^{2n-2j-2k}\,v^{2j-2r}\rangle
\end{equation}
or simplifying the notations
\begin{equation}
c_{2s,2t}=\sum_{m=0}^{s}\sum_{n=0}^{t}\frac{(-1)^{m+n}\sigma^{2m-2s}\,\tau^{2n-2t}}{2^{m+n}m!n!(2s-2m)!(2t-2n)!}\langle u^{2s-2m}\,v^{2t-2n}\rangle.
\end{equation}

\subsection{Test of the accuracy of the Gram-Charlier expansion in a 
single-subshell configuration}

\begin{figure}[htpb]
\begin{subfigure}{0.48\textwidth}
\includegraphics[width=\textwidth,angle=-90,scale=0.85]{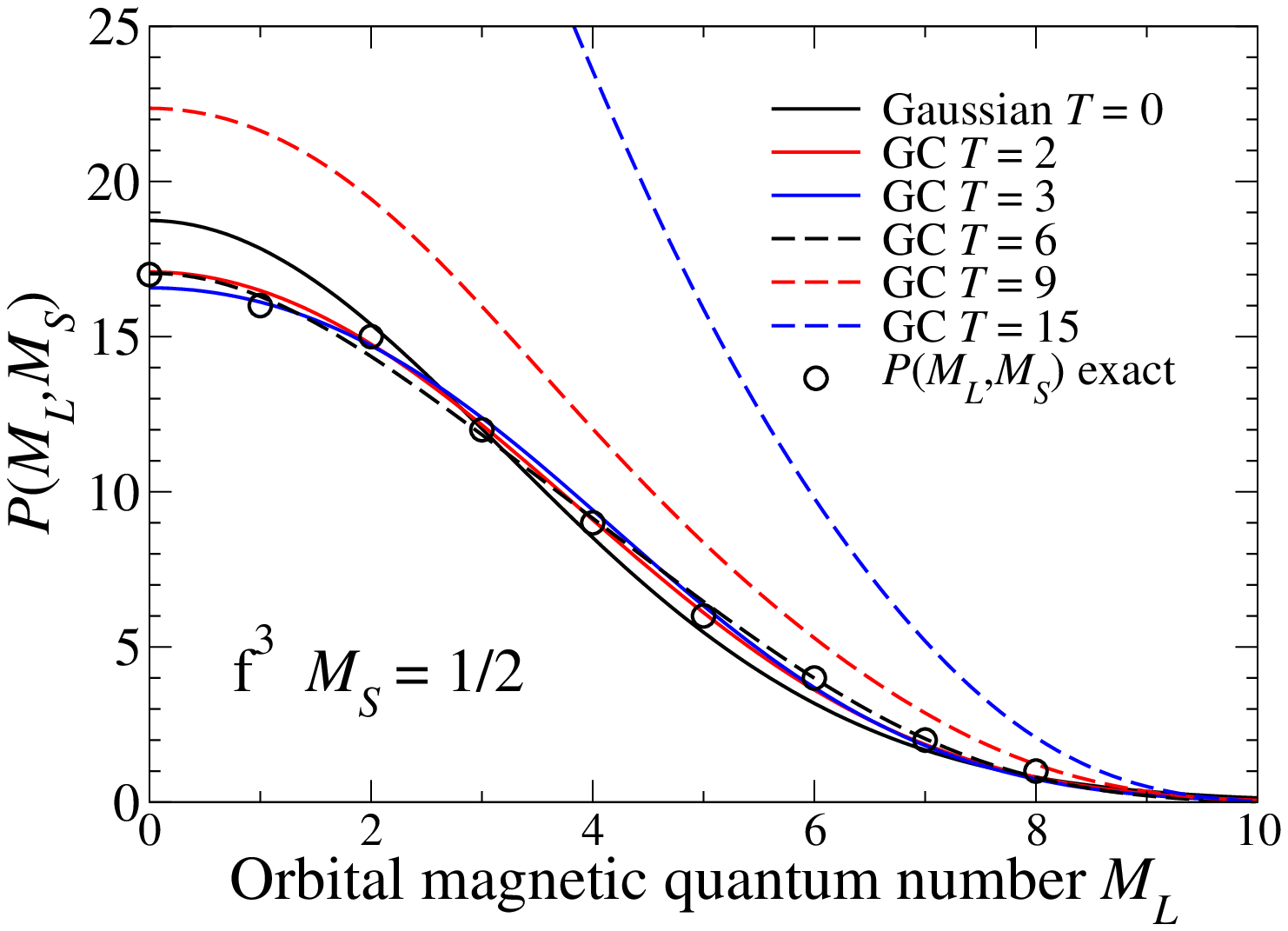}
\caption{Distribution $P(M_L,M_S=1/2)$ as a function of $M_L$. The lines 
represent the Gram-Charlier expansion for $T$=0, 2, 3, 6 and 9, 15, $2T$ being 
the order of the truncated series. Exact values exist only for integer values 
of $M_L$, $M_L\le8$.}\label{fig:GC_f3}
\end{subfigure}
\hspace{3mm}
\begin{subfigure}{0.48\textwidth}
\includegraphics[width=\textwidth,angle=-90,scale=0.85]{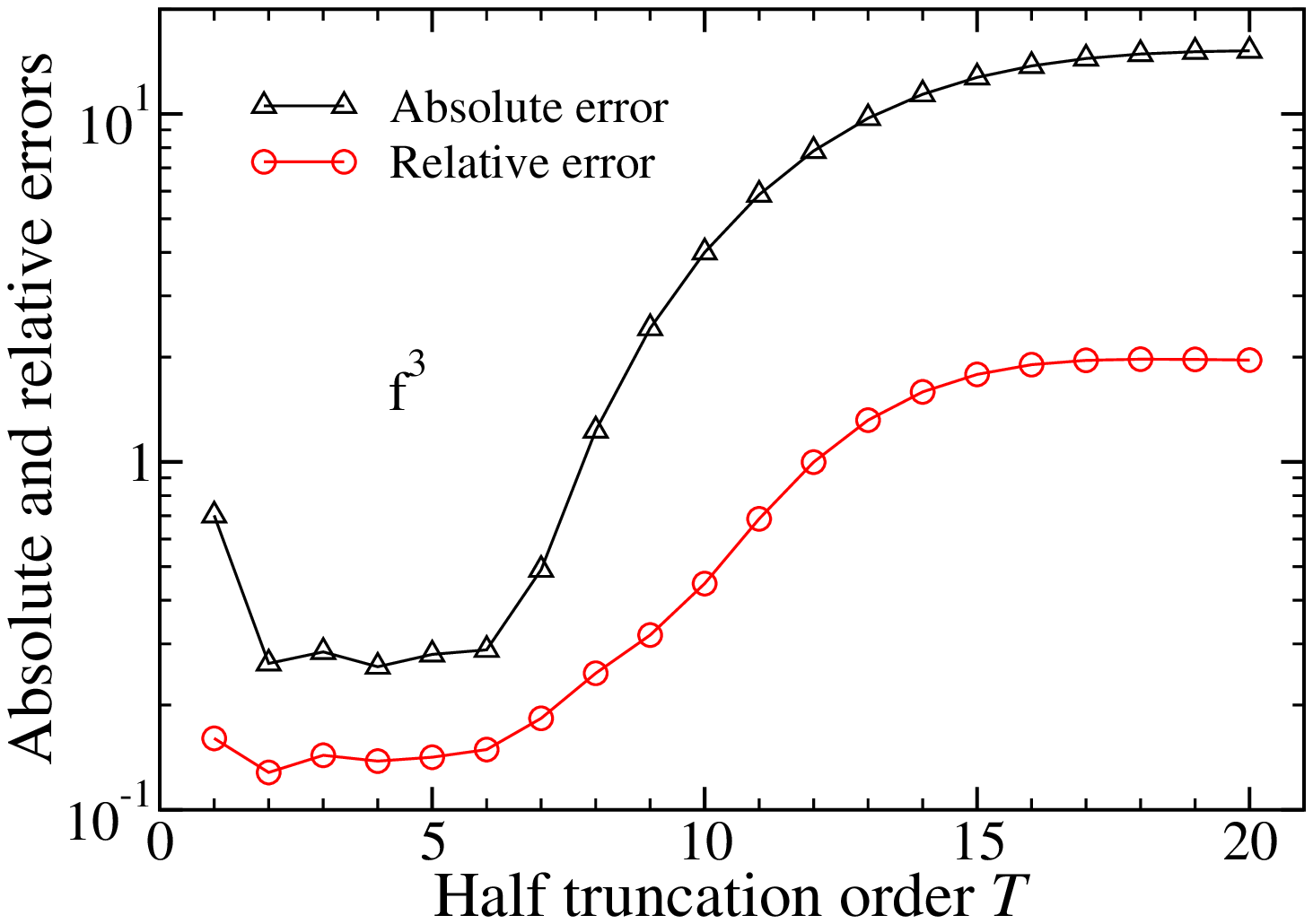}
\caption{Absolute and relative errors --- defined by Eqs.~(\ref{eq:abserr}), 
(\ref{eq:relerr}) --- on the distribution $P(M_L,M_S)$. In this plot and all 
the following error plots, the $T=1$ data points correspond to the Gaussian 
approximation.\\}\label{deltaf3}
\end{subfigure}
\caption{Gram-Charlier analysis for the configuration f$^3$.}\label{fig:f3}
\end{figure}

The Gram-Charlier expansion $\mathscr{P}_T(M_L,M_S)$ is first tested in 
the $\text{f}^3$ case in Fig.~\ref{fig:GC_f3}. The errors $\Delta_\text{abs}$ 
(\ref{eq:abserr}) and $\Delta_\text{rel}$ (\ref{eq:relerr}) are 
plotted as a function of the half truncation order $T$ in Fig.~\ref{deltaf3}.
In what follows, the ``Gaussian'' approximation is provided by the factor 
before the brackets in Eq.~(\ref{eq:GCuv}) multiplied by the configuration 
degeneracy, which is $\mathscr{N}=364$ in the f$^3$ case. Using the Gaussian 
approximation amounts to set $T=0$ or 1 in this formula, which cancels the sum 
over $n$.

We can see that the Gaussian distribution overestimates $P(M_L,M_S=1/2)$ for small values of $|M_L|$, and that a fairly accurate description is obtained for $T=2$ or 3 for all values of $M_L$. With $T=6$, the agreement is comparable to the $T=3$ case, which is confirmed by the occurrence of a plateau in the absolute and relative errors. Then the series starts to diverge (see the $T=9$ case for instance), tending asymptotically (at least in the considered range of $T$ values) to a relative error between 100 and 200 \%. 
The absolute error may be scaled by noticing that the maximum value for the 
distribution $P$ is reached if $M_L=0$, $M_S=1/2$, for which one has $P(0,1/2)=17$. 

In addition to the errors $\Delta_\text{abs}(T)$, $\Delta_\text{rel}(T)$, 
the accuracy of the Gram-Charlier representation at a given order can also be quantified by comparing the area of the latter expansion series to the quantity
$\sum_{M_L}P(M_L,M_S)$. 
Let $N_+$ be the number of fermions with spin $m_s=1/2$ and $N_-$ the number of fermions with spin $m_s=-1/2$. We have $N_+=N/2+M_S$ and $N_-=N/2-M_S$. Therefore, one obtains
\begin{equation}\label{bigG}
G_{\ell,N}(M_S)\equiv\sum_{M_L}P(M_L,M_S)=\binom{2\ell+1}{N_+}\binom{2\ell+1}{N_-}=\binom{2\ell+1}{N/2+M_S}\binom{2\ell+1}{N/2-M_S}.
\end{equation}
By simple algebraic manipulations, for instance using the generating-function formalism, we recover the Weyl-Paldus formula for the degeneracy at fixed spin $S$ \cite{PALDUS1974}:
\begin{equation}
G_{\ell,N}\left(S\right)-G_{\ell,N}\left(S+1\right)=\frac{2S+1}{2\ell+2}\binom{2\ell+2}{N/2-S}\binom{2\ell+2}{N/2+S+1}.
\end{equation}
As an example, in the $\text{f}^3$ case, we have $s_{1/2}=\sum_{M_L}P(M_L,1/2)=147$ and
$s_{3/2}=\sum_{M_L}P(M_L,3/2)=35$ and since $\sum_{M_L}P(M_L,M_S)=
\sum_{M_L}P(M_L,-M_S)$, we recover the proper total degeneracy of configuration 
$\text{f}^3$, namely $2(s_{1/2}+s_{3/2})=364=\binom{4\times 3+2}{3}.$
The quantity (\ref{bigG}) must be compared to the area of the statistical modeling by truncated Gram-Charlier expansion series (resulting from the orthogonality and normalization of Hermite polynomials):
\begin{equation}\label{integ}
\int_{-\infty}^{\infty}\mathscr{G}_T(u,v)du=\frac{e^{-v^2/2\tau^2}}{\sqrt{2\pi}\tau}\left[1+\sum_{n=2}^T c_{0,2n}\,He_{2n}(v/\tau)\right],
\end{equation}
with
\begin{equation}
    c_{0,2n}=\sum_{k=0}^n(-1)^k\frac{\tau^{2k-2n}}{2^kk!(2n-2k)!}\langle v^{2n-2k}\rangle.
\end{equation}
Table \ref{tab8} displays the area of the Gram-Charlier series (integrated between $M_L=-\infty$ and $\infty$) for different orders $2T$ in the cases $M_S=1/2$ and $M_S=3/2$. The right-hand side of Eq.~(\ref{integ}) is multiplied by the total degeneracy $\binom{14}{3}=364$. 

\begin{table}[htbp]
\renewcommand{\arraystretch}{1.5}
\centering
\begin{tabular}{@{\quad}c@{\quad}*{9}{c}}\hline\hline
$2T$ & 2 & 4 & 6 & 8 & 10 & 12 & 14 & 16 \\\hline
$M_S=1/2$ & 149.697 & 147.439 & 147.485 & 146.606 & 145.466 & 147.342 & 155.585 & 172.071 \\
$M_S=3/2$ & 30.9650 & 34.3295 & 34.3410 & 35.3613 & 36.5085 & 35.1317 & 31.5573 & 29.0911 \\\hline\hline
\end{tabular}
\caption{Area of the Gram-Charlier series (integrated between $M_L=-\infty$ and $\infty$) in the case of the $\text{f}^3$ configuration 
for different orders $2T$ and $M_S=1/2$ (second line) as well as $M_S=3/2$ (third line). The values are obtained taking the right-hand side of Eq. (\ref{integ}) multiplied by the total degeneracy of $\text{f}^3$, which is equal to 364. The exact value of $\sum_{M_L}P(M_L,1/2)$ is 147 and the exact value of $\sum_{M_L}P(M_L,3/2)$ is 35.}\label{tab8}
\end{table}

\begin{figure}[htbp]
\begin{subfigure}{0.48\textwidth}
\includegraphics[width=\textwidth,angle=-90,scale=0.85]{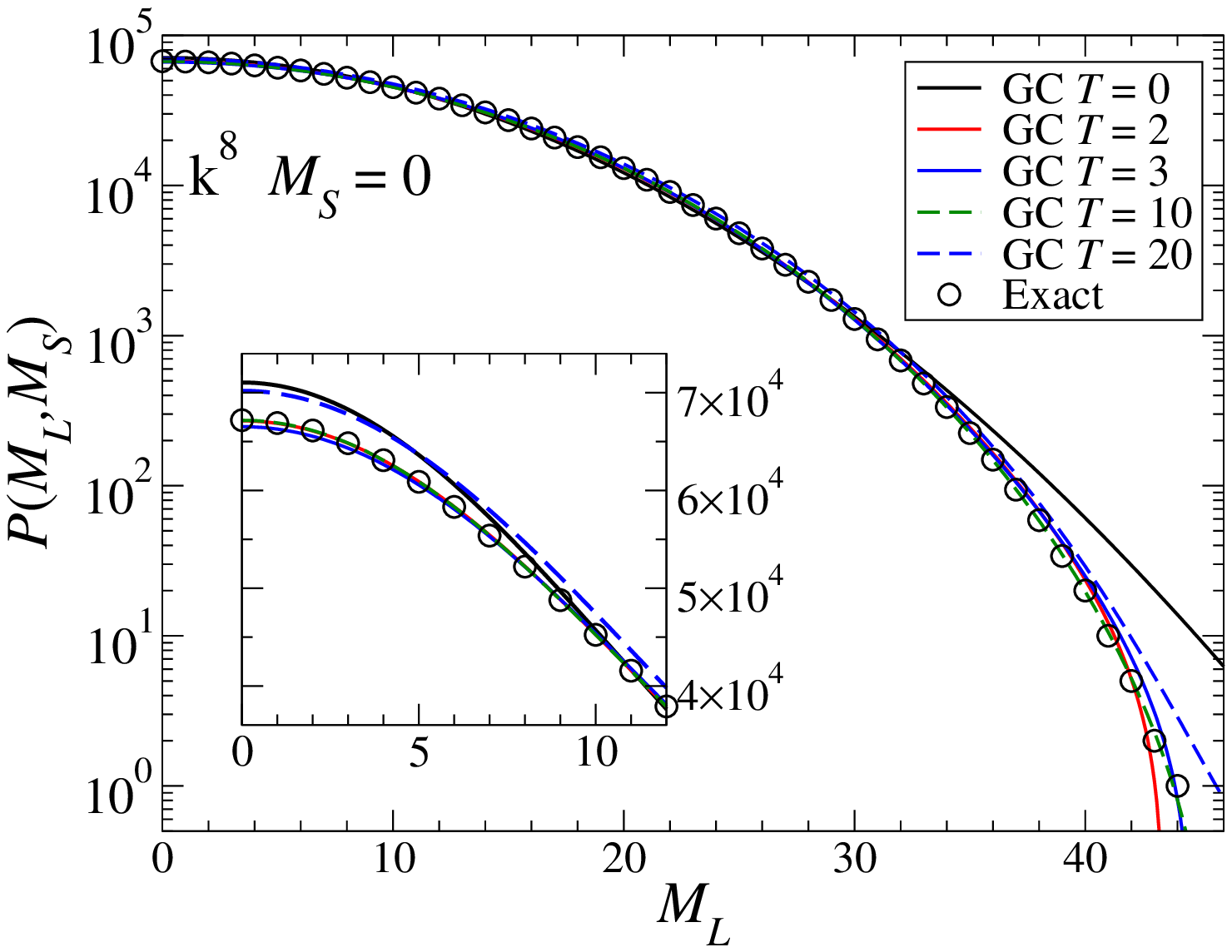}
\caption{Gram-Charlier approximation and exact values of 
  $P(M_L,M_S=0)$ in semi-log coordinates. 
  The inset shows an enlarged view of the $M_L\le12$ region, in 
  linear coordinates. In this region the curves labeled $T=2$ and 10 
  are indistinguishable at the drawing accuracy.}\label{fig:GC_k8}
\end{subfigure}
\hspace{3mm}
\begin{subfigure}{0.48\textwidth}
\includegraphics[width=\textwidth,angle=-90,scale=0.85]{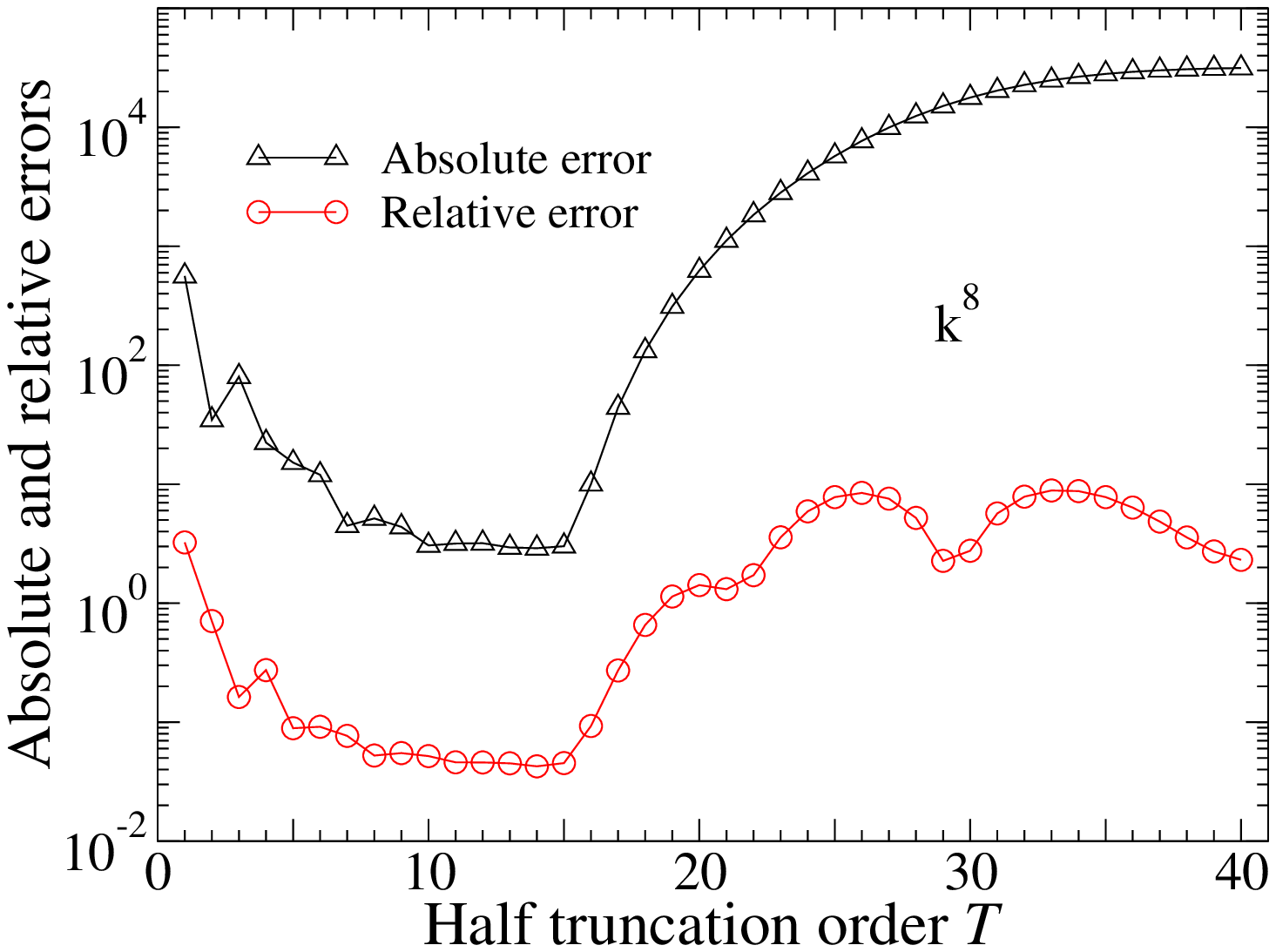}
\caption{Error defined by Eq.~(\ref{eq:abserr},\ref{eq:relerr}) on the 
distribution $P(M_L,M_S)$. The absolute error (\ref{eq:abserr}) may be 
scaled by the maximum value $P(M_L=0,M_S=0)=67189$. As above the $T=1$ 
value is by convention the Gaussian approximation.}\label{fig:err_k8}
\end{subfigure}
\caption{Gram-Charlier analysis for the configuration k$^8$.}
\label{figs:GC_k8}
\end{figure}

In order to demonstrate how the approximation improves for a greater 
number of electrons, we have plotted in Figs.~\ref{fig:GC_k8} and 
\ref{fig:err_k8} the corresponding Gram-Charlier data for the configuration 
k$^8$ ($\ell=7$). Looking at the $M_S=0$ data in Fig.~\ref{fig:GC_k8}, 
we notice that 
the lowest order $T=0$, i.e., the simple Gaussian 
factor in Eq.~(\ref{eq:GCuv}), provides an acceptable approximation only 
if $10<M_L<35$. The $T=2$ approximation is correct over the whole 
range, except for $M_L\ge43$. The $T=3$ approximation performs quite well 
over the whole $M_L$ range (though being one per-cent too low for $M_L=0$, 
while $T=2$ and 4 are then accurate at the per-thousand level).
When $T$ increases up to 15, the distribution is almost $T$-independent: 
the $T=10$ and 15 data would be indistinguishable at Fig.~\ref{fig:GC_k8} 
drawing accuracy. For greater $T$ values the approximation quality 
deteriorates for any $M_L$ value as seen on the $T=20$ curve. 
Looking into more detail, the absolute and relative errors plotted in 
Fig.~\ref{fig:err_k8} significantly decreases for $T<10$, while the 
relative error was almost constant for $1<T<6$ in the $\text{f}^3$ case.
For $T>15$, as seen on Fig.~\ref{fig:err_k8}, the series begins 
to diverge, and including more terms may result in a poorer approximation. 
This observation is similar to the one made for single-variable 
Gram-Charlier series \cite{POIRIER2021a}.

\subsection{Accuracy of the Gram-Charlier expansion in a 
multiple-subshell configuration}

As is well known the interest of cumulants is that they are additive, 
e.g., the cumulants of the configuration p$^2$d$^3$ are the sums of the 
cumulants of p$^2$ and d$^3$ configurations. One may then use the 
above mentioned values of single-subshell cumulants. 

\begin{figure}[htbp]
\begin{subfigure}{0.48\textwidth}
\includegraphics[width=\textwidth,angle=-90,scale=0.85]{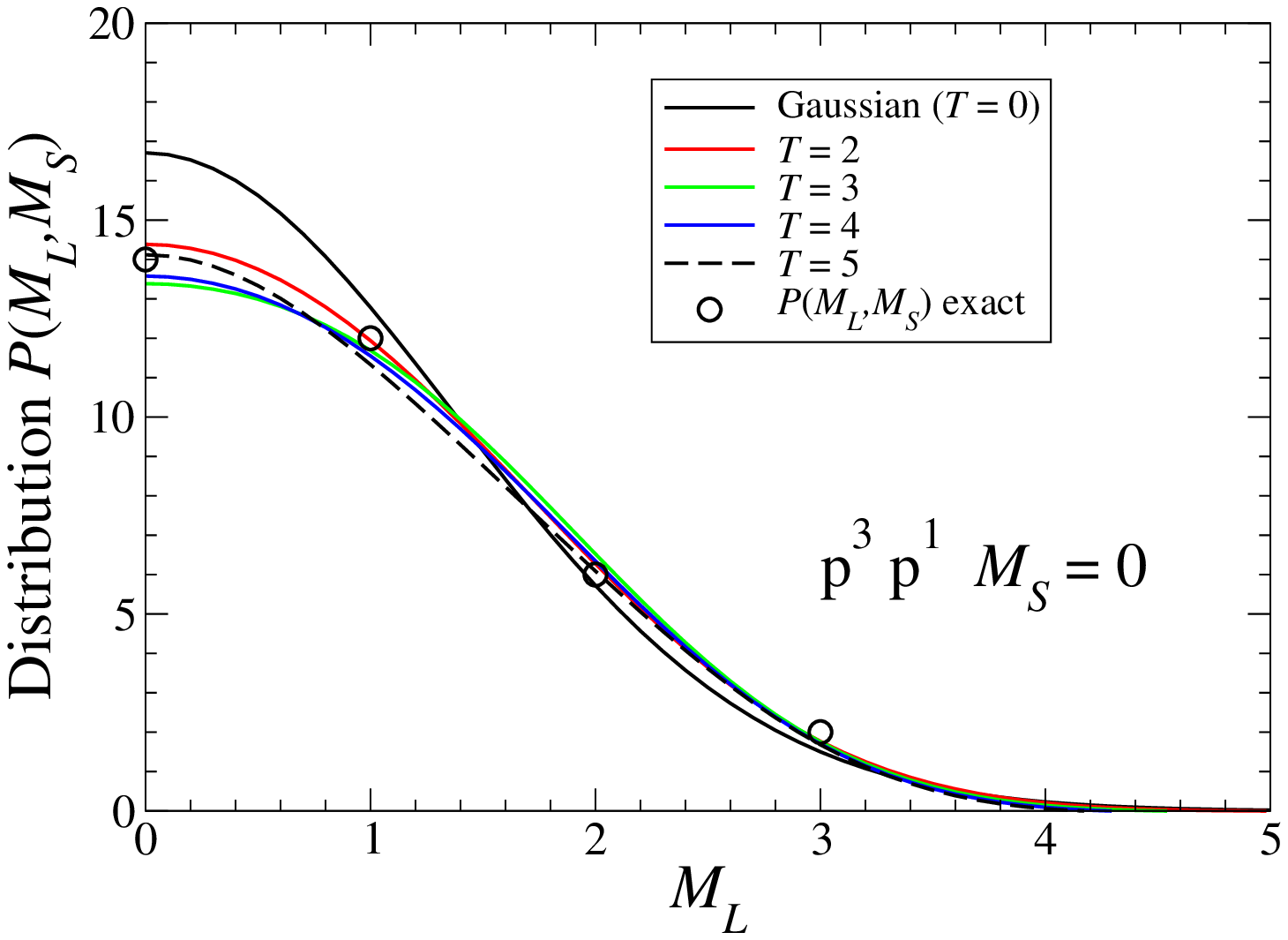}
\caption{Gram-Charlier approximations (solid lines) and exact values 
(circles) of the distribution $P(M_L,M_S=0)$.\\\\}\label{fig:GCp3p1}
\end{subfigure}
\hspace{3mm}
\begin{subfigure}{0.48\textwidth}
\includegraphics[width=\textwidth,angle=-90,scale=0.85]{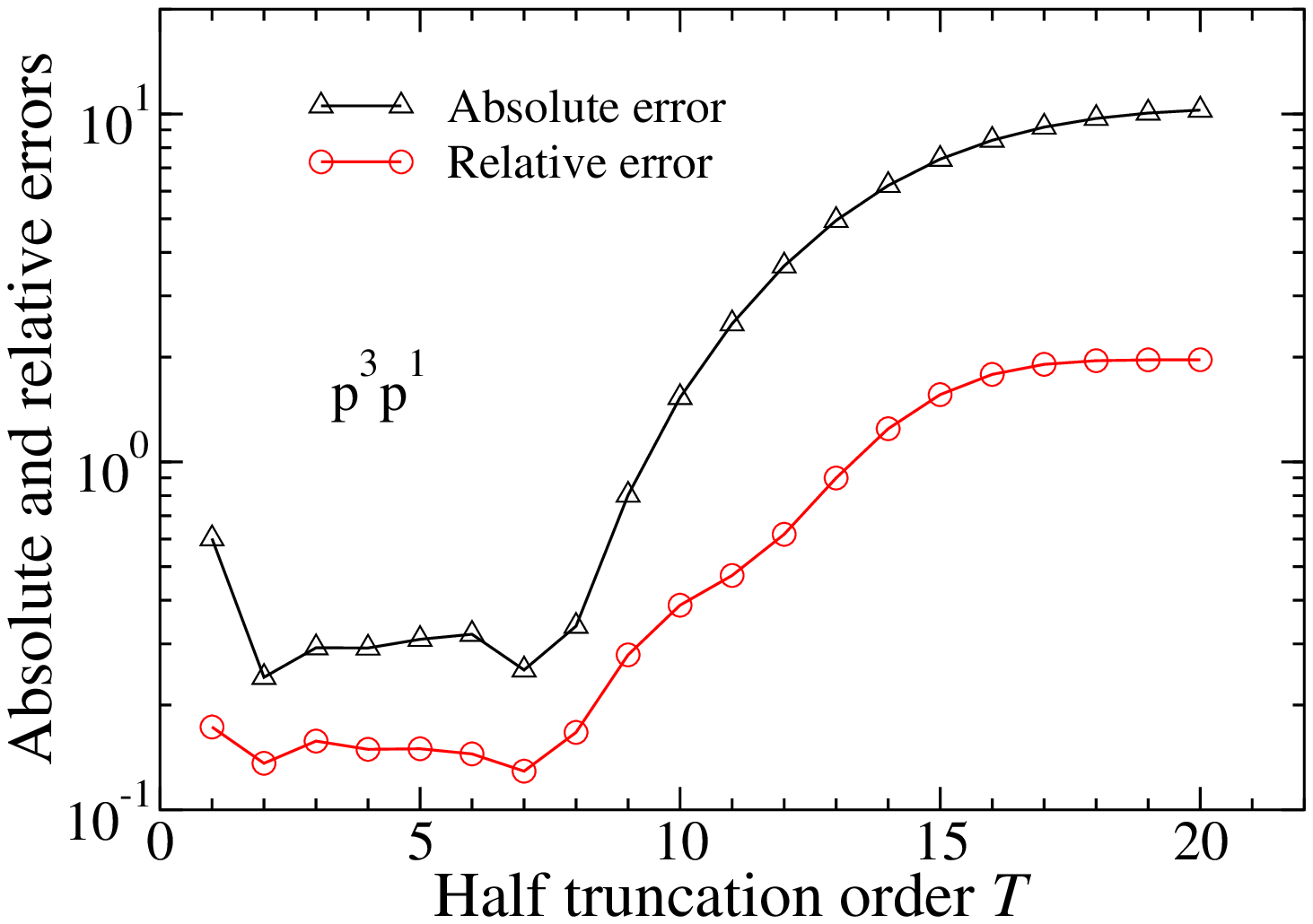}
\caption{Error defined by Eqs.~(\ref{eq:abserr}), (\ref{eq:relerr}) 
on the distribution $P(M_L,M_S=0)$. The absolute error (\ref{eq:abserr}) 
may be scaled knowing that $P(M_L=0,M_S=0)=14$. The $T=1$ data correspond 
to the Gaussian case.}\label{fig:deltap3p1}
\end{subfigure}
\caption{Gram-Charlier analysis for the 2-subshell configuration p$^3$p$^1$.} 
\label{figs:GC_p3p1}
\end{figure}

As a first example of multiple-subshell configuration, we present on Fig. \ref{fig:GCp3p1} the Gram-Charlier 
approximation of $P(M_L,M_S=0)$ in the case of the configuration $\text{p}^3 \text{p}^1$, 
i.e., containing 3 equivalent p electrons and another p electron on a 
distinct subshell, as in $2\text{p}^3 3\text{p}$. The curve 
labeled as $T=0$ corresponds to the plain Gaussian bi-variate function, 
as given by 
Eq.~(\ref{eq:GCuvN}) where $\mathscr{G}_T$ is replaced 
by the factor in front of the brackets in Eq.~(\ref{eq:GCuv}).
The curves labeled as 
$T=2,3,4,5$ correspond to the various corrections appearing in the series. 
In this case, we note that the $T=2$ correction improves the quality 
of the distribution, while taking into account higher orders does not 
bring significant changes. 
This plot displays clear similarities with the $\text{f}^3$ case (Fig.~\ref{fig:GC_f3}). 
Comparing the absolute errors in Figs.~\ref{deltaf3} and \ref{fig:deltap3p1}, 
we note that in the both cases the plateau $T=2$--7 is about 3 times 
below the $T=1$ (Gaussian) value. The relative error levels off at about 
0.15 for $T<8$ values, while it increases up to $\Delta_\text{rel}\simeq2$ 
for large $T$. 

\begin{figure}[htbp]
\begin{subfigure}{0.48\textwidth}
\includegraphics[width=\textwidth,angle=-90,scale=0.85]{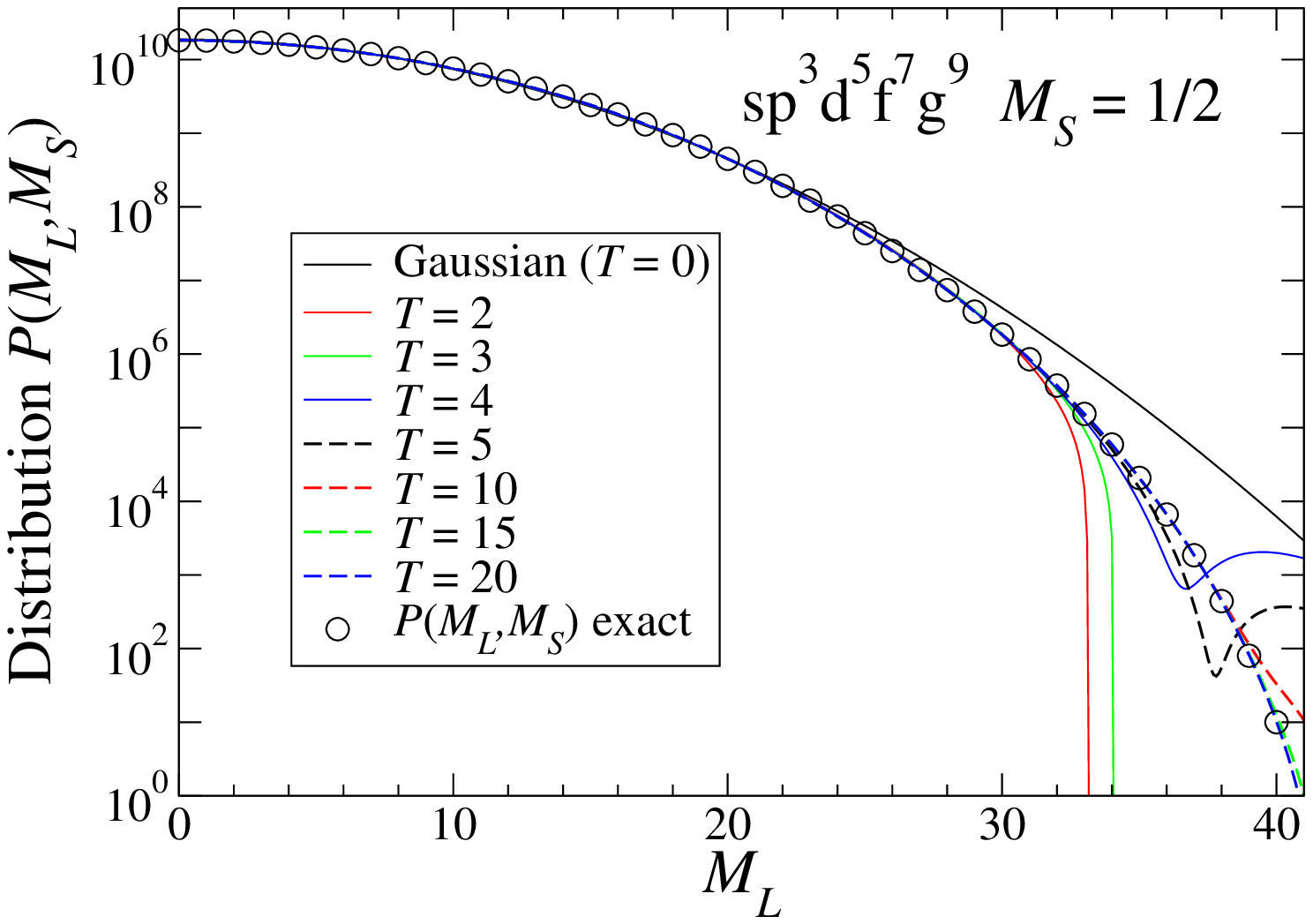}
\caption{Gram-Charlier approximations (lines) and exact values 
(circles) of the distribution $P(M_L,M_S=1/2)$. The various curves correspond 
to various half truncation orders $T$ in the Gram-Charlier series.}\label{fig:GCsp3d5f7g9}
\end{subfigure}
\hspace{3mm}
\begin{subfigure}{0.48\textwidth}
\includegraphics[width=\textwidth,angle=-90,scale=0.85]{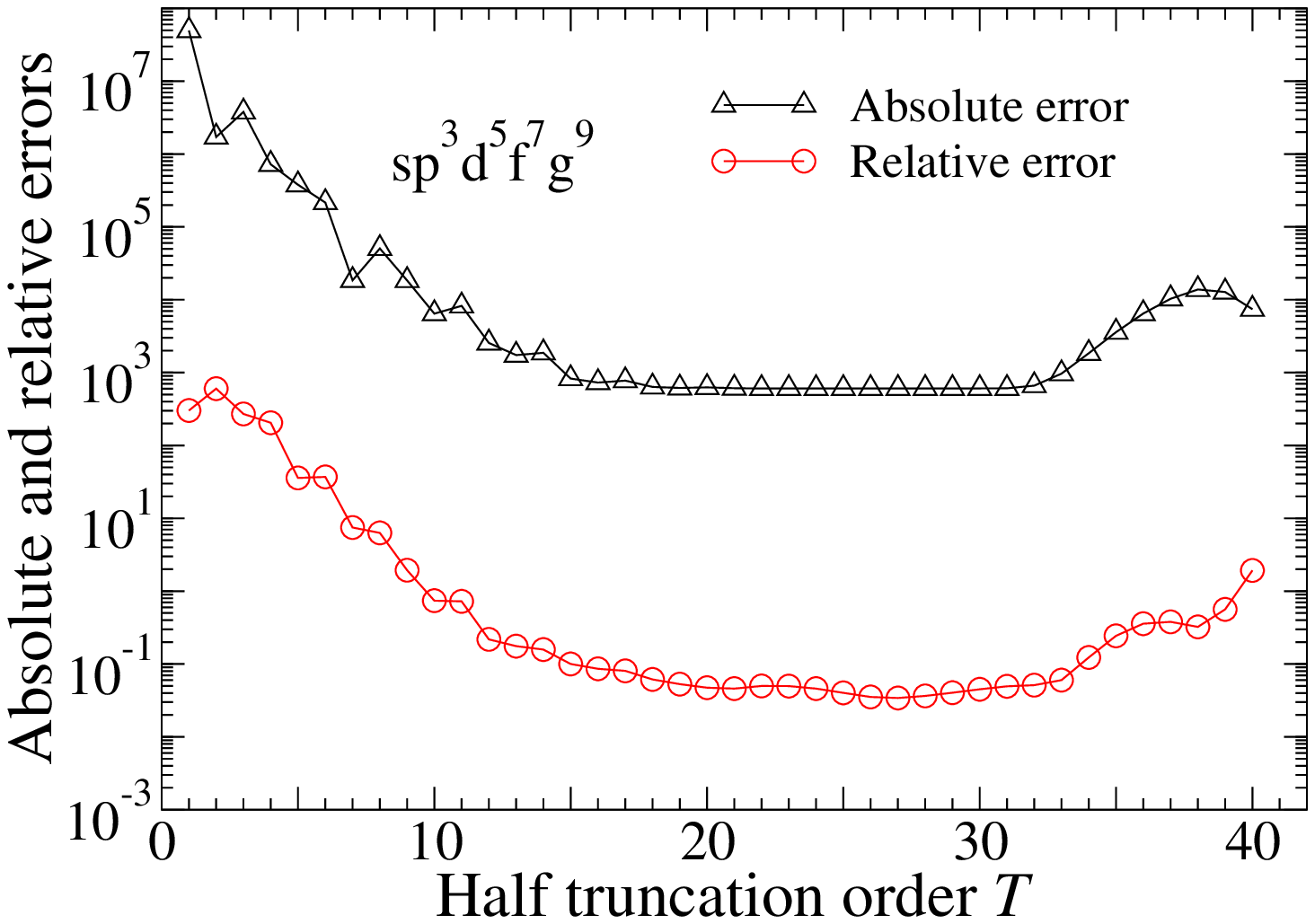}
\caption{Absolute and relative errors defined by Eqs.~(\ref{eq:abserr},
\ref{eq:relerr}) of the distribution $P(M_L,M_S=1/2)$. The absolute error 
may be scaled noting that $P(M_L=0,M_S=1/2)=1.8297912590\times10^{10}$.}\label{fig:err_GC_5sub}
\end{subfigure}
\caption{Gram-Charlier analysis for the configuration sp$^3$d$^5$f$^7$g$^9$.}
\label{figs:GCsp3d5f7g9}
\end{figure}

Since it is useful to consider configurations with a greater number of 
electrons we show on Fig. \ref{fig:GCsp3d5f7g9} the Gram-Charlier analysis 
in the 5-subshell case sp$^3$d$^5$f$^7$g$^9$. One observes 
that for this 25-electron configuration, the Gram-Charlier series performs 
more efficiently than in the 4-electron case p$^3$p$^1$ considered above. 
Noticeably, the $T=3$ (respectively 4) truncation provides 
a poor approximation of $P(M_L,M_S)$ if $M_L\ge33$ 
(respectively 34) while greater $T$-values lead to 
a more acceptable representation. For instance the $T=10$ and 15 
curves are almost superimposed at the drawing accuracy, but the latter 
approximation is better for the maximum value $M_L=40$. One has then 
$P(40,1/2)=10$, to be compared to 32.94, 12.46, and 10.16 for $T=10$, 
15, and 20 respectively.
Looking at absolute and relative errors in Fig.~\ref{fig:err_GC_5sub}, 
we observe that increasing $T$ from 1 to 15 results in an improvement in the 
Gram-Charlier approximation by 4 orders of magnitude. This is even better 
than in the k$^8$ case (Fig.~\ref{fig:err_k8}) where the error was only 
lowered by a factor of $\sim100$. As mentioned in the caption, when 
assessing the quality of the present approximation, one must compare the 
absolute error with a plateau at about 600 to the maximum value 
$P(M_L,M_S)<1.83\times10^{10}$. In the present case for $T>32$ the series 
begins to diverge, though we did not explore its behavior for very large $T$. Finally one may notice that one has $\Delta_\text{rel}(T)>0.03$ even 
in the most favorable case $T\sim20$, which is due to the approximation 
for $P(M_L=40,M_S)$ which is of moderate quality.

\begin{figure}[htbp]
\begin{subfigure}{0.48\textwidth}
\includegraphics[width=\textwidth,angle=-90,scale=0.85]{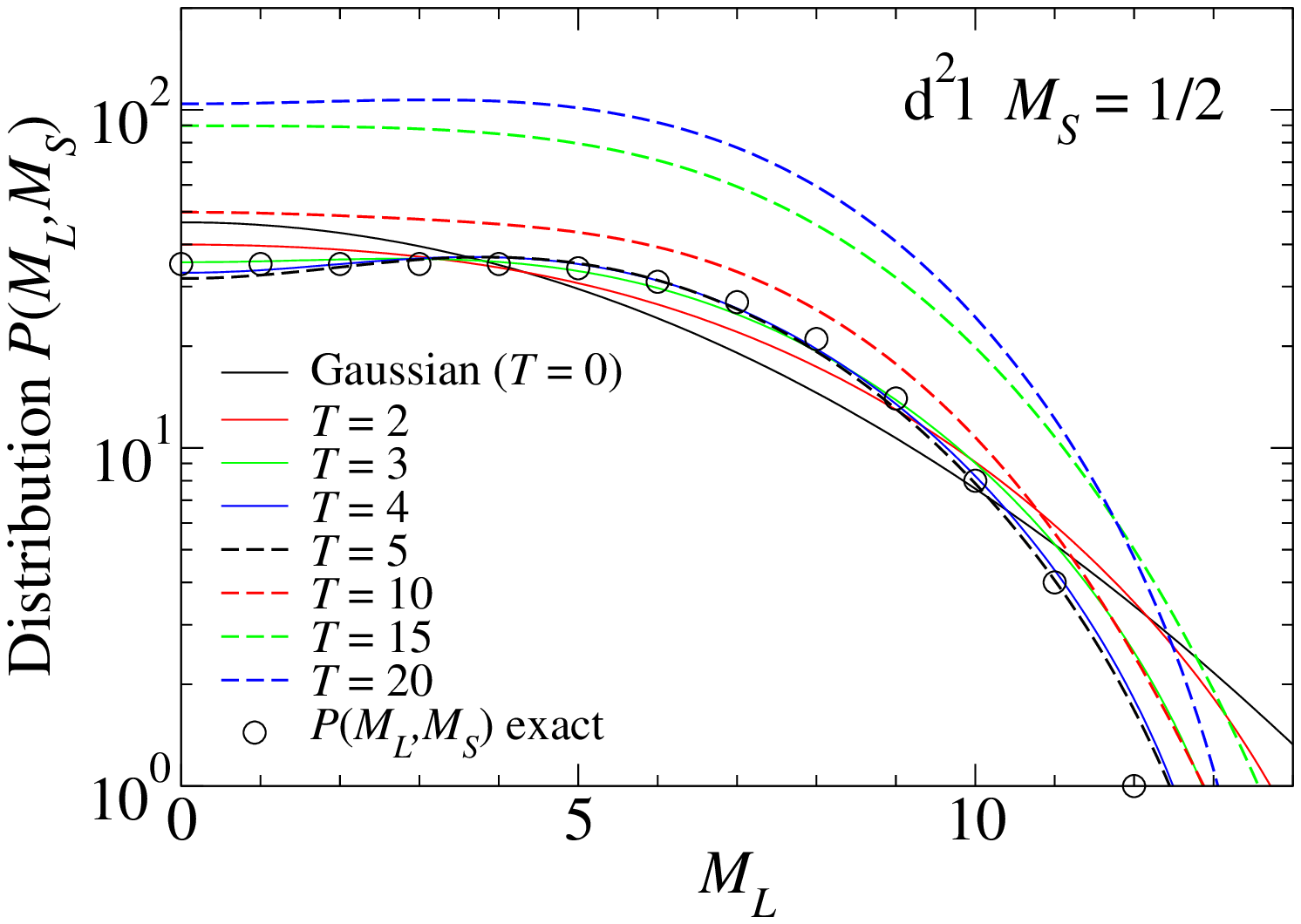}
\caption{Gram-Charlier approximations (lines) and exact values 
(circles) of the distribution $P(M_L,M_S=1/2)$.\\\\}\label{fig:GCd2l}
\end{subfigure}
\hspace{3mm}
\begin{subfigure}{0.48\textwidth}
\includegraphics[width=\textwidth,angle=-90,scale=0.85]{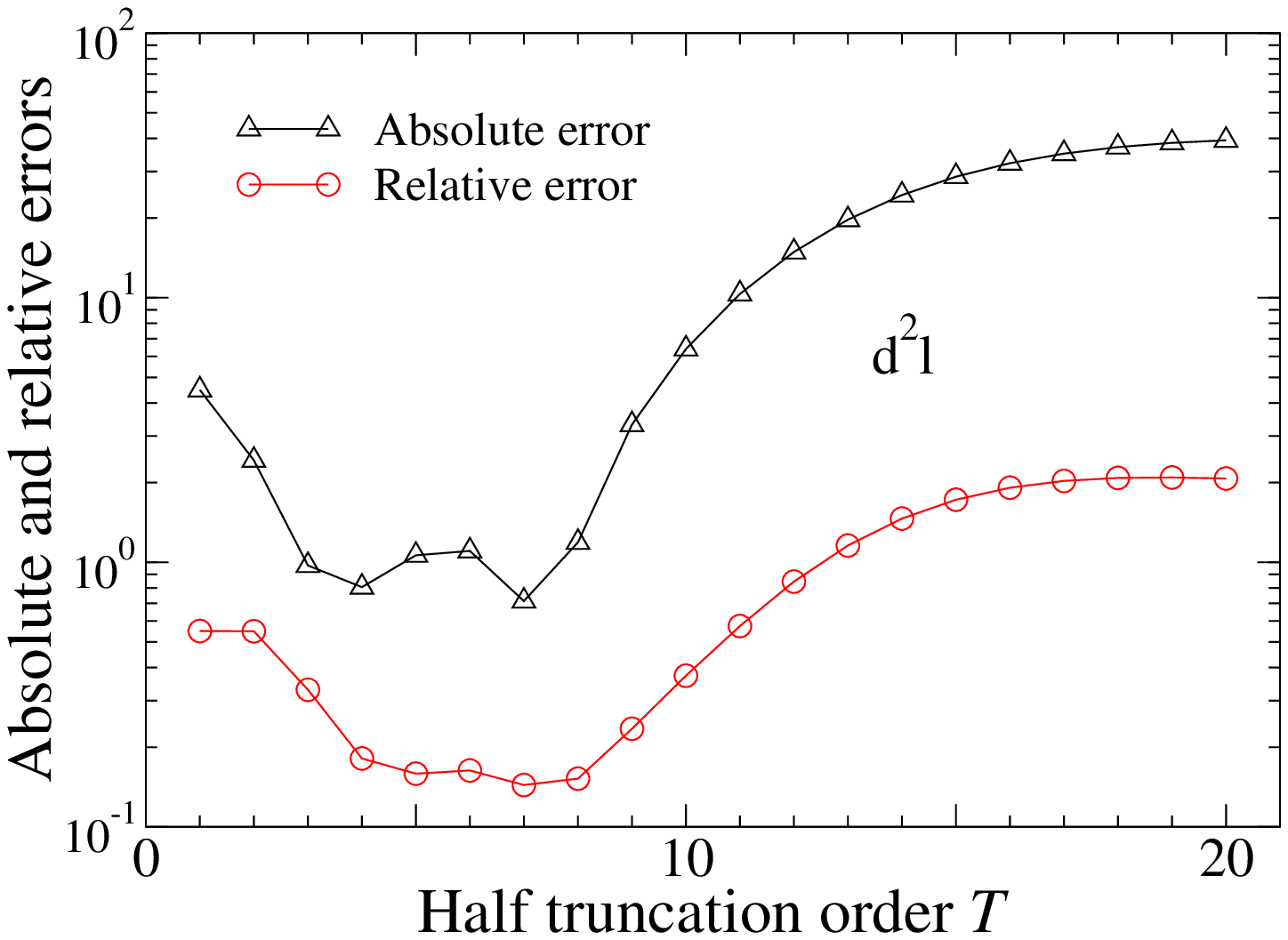}
\caption{Absolute and relative errors defined by Eqs.~(\ref{eq:abserr}),
(\ref{eq:relerr}) on the distribution $P(M_L,M_S)$. The absolute error may 
be scaled noting that $P(M_L,M_S=1/2)=35$ if $|M_L|\le4$.}\label{fig:err_d2l}
\end{subfigure}
\caption{Gram-Charlier analysis for the configuration d$^2$l (two electrons 
$\ell=2$, one electron $\ell=8$).}\label{figs:GC_d2l}
\end{figure}

As a last example, we may analyze the two-subshell case d$^2$l, 
i.e., with two $\ell=2$ electrons and a single $\ell=8$ electron. 
As noticed before \cite{GILLERON2009}, when a configuration contains 
several low-$\ell$ electrons together with a large-$\ell$ electron, 
the momentum distribution exhibits a wide plateau in its center. This 
case is analyzed in Figs. \ref{fig:GCd2l} and \ref{fig:err_d2l}. 
As expected, one observes that the Gaussian approximation $T=1$ is a poor 
representation of the exact $P(M_L,M_S=1/2)$ distribution. Nevertheless 
higher-order values such as $T\simeq5$ provide an acceptable approximation 
of the whole-range distribution. Looking at the absolute and relative errors 
in Fig.~\ref{fig:err_d2l}, we notice that, as in the previous cases with 
few electrons ($\text{f}^3$, p$^3$p$^1$), the $T\simeq5$ expansion improves the 
simple Gaussian value by less than an order of magnitude. However in the 
present case, increasing $T$ from 1 to 5 results in a decrease of 
$\Delta_\text{rel}$ by a factor greater than 3, 
while the improvement was mostly 
negligible in the the f$^3$ case or in the p$^3$p$^1$ case, as seen on 
Fig.~\ref{deltaf3} or Fig.~\ref{fig:deltap3p1}).

\section{Conclusion}\label{sec7}

In this paper we have shown that the distribution of $(M_L,M_S)$ magnetic quantum numbers, from which the number of $LS$ spectroscopic terms $Q(L,S)$ is deduced, can be obtained numerically 
using an efficient recurrence relation generalizing the one described in 
Ref.~\cite{GILLERON2009} for $J$ and $M_J$. 
A recurrence relation for the two-variable generating function of the cumulants of the ($M_L$, $M_S$) joint distribution was derived, together with an explicit form involving simple partitions. The relation was also extended to the case where the second angular momentum (spin in atomic physics, isospin in nuclear physics) is not necessarily equal to 1/2 (i.e., for two arbitrary angular momenta). Based on an analogy with the Newton-Girard identities for elementary symmetric polynomials, an explicit formula for the cumulant generating function was also provided. 
A bi-variate Gram-Charlier expansion has been proposed to provide an 
analytical approximation for the $P(M_L,M_S)$ distribution. It has been 
checked that such expansion with few terms included is particularly 
efficient for configurations with a large number of electrons. Such series 
with less than a dozen terms may even represent the momentum distribution 
even for configuration with few electrons in an acceptable way, even 
in special cases where the $P(M_L,M_S)$ distribution exhibits a wide plateau. 
Nevertheless as for the single-variable $P(M_J)$ case studied in previous 
works, it turns out that the Gram-Charlier series diverges, with an onset of 
this divergence for a number of terms increasing with the total number of 
electrons in the configuration.
The Gram-Charlier-series modeling of $P(M_L,M_S)$ makes it possible to estimate quite accurately the number of lines between two non-relativistic configurations. Such a number is important for the computation of hot-plasma radiative opacity, in particular when combining statistical methods \cite{BAUCHE1988} and fine-structure calculations.

\appendix

\section{Recurrence over the number of fermions for the cumulant generating function: case of a general spin}\label{appA}

The recurrence relation for generating functions can be rewritten with minor changes in the case where the fermions are characterized by two moments $j_u,j_v$, 
the case of the present work corresponding to $j_u=\ell$ and $j_v=1/2$. This seems rather formal in an atomic-physics framework, but enables one to deal with the 3/2 isospin case, relevant in particle physics (see for instance Ref. \cite{LANG2012}). 
Indeed  for any ``orbital momentum'' $j_u$ and ``spin momentum'' $j_v$, the 
\emph{single-particle} generating function for arguments $(u,v)$ is easy to 
obtain, as shown in this Appendix. 

To each fermion is attributed a pair of indices $(\mu_i,\nu_i)$ where $\mu_i$ is the ``orbital'' (respectively spin-orbital) magnetic quantum number and $\nu_i$ the ``generalized-spin'' magnetic quantum number 
(respectively the isospin projection) of fermion $i$ in the atomic-physics (respectively nuclear-physics) case. The composite index becomes
\begin{equation}
\xi_i=(2j_v+1)\mu_i+\nu_i+2j_uj_v+j_u+j_v+1
\text{ with } -j_u\le\mu_i\le j_u \text{ and } -j_v\le\nu_i\le j_v,
\end{equation}
where $j_u$ denotes the ``orbital momentum'' and $j_v$ the ``spin''. The index $\xi_i$ varies from 1 to 
$(2j_u+1)(2j_v+1)$. The proof is similar to the 1/2-spin case. 
Because of the Pauli exclusion principle, all the $\xi_i$ coefficients of fermions of a configuration (respectively subshell) are distinct. 

The two-variable generating function $s(N,j_u,j_v,u,v)$ --- 
we add an additional argument to the $s$ function considered in Sec.~\ref{sec3}
since two moments are involved now ---
can be put in a compact form relating the $N=1$ case to the 
$N>1$ case. This can be derived in a similar way as Eq.~(\ref{eq:recNfngen}).
One obtains
\begin{equation}\label{eq:recSN2mts}
s(N,j_u,j_v,u,v)=\frac1N
\sum_{p=1}^{N}(-1)^{p+1}s(N-p,j_u,j_v,u,v)s(1,j_u,j_v,pu,pv).
\end{equation}
The one-fermion generating function can be obtained from the generalization of Eq. (\ref{eq:vals1}):
\begin{align}
s(1,j_u,j_v,u,v)&=\sum_{|\mu_1|\le j_u,|\nu_1|\le j_v} e^{\mu_1u+\nu_1v}=\sum_{\mu_1}e^{\mu_1u}\sum_{\nu_1}e^{\nu_1v}=\frac{\sinh\left[(j_u+1/2)u\right]}{\sinh(u/2)}\cdot\frac{\sinh\left[(j_v+1/2)v\right]}{\sinh(v/2)}\label{eq:s1an2mts}
\end{align}
and the second factor of Eq. (\ref{eq:s1an2mts}) is, as expected, equal to $2\cosh(v/2)$ if $j_v=1/2$.

The single-momentum case is recovered setting $j_v=0$ which shows 
that the formula remains valid for a single momentum $j=j_u$.

\section{Newton-Girard identities and explicit formula for the cumulant generating function}\label{appB}

The cumulant generating function reads
\begin{equation}
s(k,\ell,u,v)=\sum_{1\le i_1<i_2<\cdots<i_k\le N} 
X_{i_1}X_{i_2}\cdots X_{i_k},
\end{equation}
with $X_i=e^{(m_{\ell,i}u+m_{s,i}v)}$ and $s(0,\ell,u,v)=1$. Let us calculate the generating function
\begin{equation}\label{eq:defXi}
\Xi=\sum_{0\le k\le\infty} s(k,\ell,u,v)\, t^k,
\end{equation}
noting that in fact the latter summation is finite, since $s(k,\ell,u,v)=0$ if $k>n$. One has, identifying the coefficients of the $t^k$ terms:
\begin{equation}\label{eq:Xiprod}
\Xi=\prod_{1\le i\le N}(1+X_it).
\end{equation}
Indeed, in the products of the $N$ factors above, the $t^k$ term is the sum of the $\binom{N}{k}$ 
products of $k$ factors $X_{i_1}X_{i_2}\cdots X_{i_k}$, with $1\le i_1<i_2<\cdots<i_k\le N$. Let us rewrite (\ref{eq:Xiprod}) in the form
\begin{equation}
\Xi=\prod_{1\le i\le N}\exp\left[\log(1+X_it)\right]=\exp\left[\sum_{1\le i\le N}\log(1+X_it)\right].
\end{equation}
With the expansion $\log(1+x)=x-x^2/2+x^3/3+\cdots\text{ 
convergent if }|x|<1$,
one gets, if $|\max(X_i)t|<1$ (for analyticity reasons, the expansion remains actually valid whatever the modulus of $t$),
\begin{align}
\Xi&= \exp\left(\sum_{1\le i\le n}\sum_{s=1}^\infty
\frac{(-1)^{s-1}}{s} X_i^st^s\right)
= \exp\left(\sum_{s=1}^\infty (-1)^{s-1}\frac{\mathscr{S}_s}{s}t^s\right)=\sum_{m=0}^\infty \frac{1}{m!}\left(\sum_{s=1}^\infty (-1)^{s-1}\frac{\mathscr{S}_s}{s}t^s\right)^m,\label{eq:Xiexp}
\end{align}
with
\begin{equation}
\mathscr{S}_k=\sum_{1\le i\le N} X_i^k.
\end{equation} 
The $m-$th power in the sum may be computed with the 
identity used in our previous work \cite{PAIN2020} (see also section 24.1.2 in Ref. \cite{ABRAMOWITZ1972}):
\begin{equation}
\frac{1}{m!}\left(\sum_{N=1}^\infty x_n\frac{t^N}{N!}\right)^m = 
\sum_{N=m}^\infty t^N\sum_{a_1,a_2,\cdots, a_N}
\prod_{1\le j\le N}\frac{(x_j/j!)^{a_j}}{a_j!}
\end{equation}
involving the partition number $\mathscr{P}(N;a_1,\cdots, a_N)$, 
and where integer indices $a_1,a_2,\cdots, a_N$ are constrained by
\begin{align}
a_1+a_2+\cdots+a_N&=m\nonumber\\
a_1+2a_2+\cdots+Na_N&=N.\label{eq:sum_iai}
\end{align}
With the substitution $x_j/j!\rightarrow (-1)^{j-1}\mathscr{S}_j/j$, the new expression of the expansion of the generating function 
$\Xi$ (see Eq. (\ref{eq:Xiexp})) reads
\begin{equation}
\Xi = \sum_{m=0}^\infty\sum_{N=m}^\infty t^N
\sum_{a_1,a_2,\cdots, a_N}\prod_{j=1}^N\frac{\left[(-1)^{j-1}\mathscr{S}_j/j\right]^{a_j}}{a_j!}
\end{equation}
where the $a_i$ satisfy the constraints (\ref{eq:sum_iai}). In the above expression, the double sum over $m,N$ 
can be replaced by a single sum over $N$, and it remains
\begin{equation}
\Xi = \sum_{N=0}^\infty (-1)^N t^N
\sum_{a_1,a_2,\cdots, a_N}\prod_{j=1}^N\frac{\left(-\mathscr{S}_j/j\right)^{a_j}}{a_j!},
\end{equation}
where because of constraint (\ref{eq:sum_iai}), we replaced the product of factors $(-1)^{ja_j}$ by $(-1)^N$. This gives
\begin{equation}\label{eq:relekpj}
s(N,\ell,u,v)=(-1)^N \sum_{\stackrel{a_1,a_2,\cdots, a_N}{a_1+2a_2+\cdots +Na_N=N}} \prod_{j=1}^N \frac{\left(-\mathscr{S}_j/j\right)^{a_j}}{a_j!}.
\end{equation}
A similar proof was given by Richter \cite{RICHTER1950}. Setting 
\begin{equation}
f(t)=\sum_{k=0}^N s(k,\ell,u,v)t^k,\quad g(t)=\sum_{k=0}^\infty \mathscr{S}_kt^k,\text{\quad with }
f(0)=N,\quad g(0)=\mathscr{S}_0=N,
\end{equation}
one can check easily that $g(t)=N-tf'(t)/f(t)$. Indeed, as we have seen,
$f(t)=\prod_{j=1}^N(1+X_jt)$, and thus the logarithmic derivative can be reformulated with the Taylor expansion of $1/(1+X)$, 
\begin{equation}
\frac{f'(t)}{f(t)}=\sum_j\frac{X_j}{1+X_jt}=
\sum_{j=1}^N X_j\sum_{k=0}^\infty(-1)^kX_j^kt^k
=-\frac{1}{t}\sum_{k=0}^\infty(-1)^{k+1}\mathscr{S}_{k+1} t^{k+1}=-\frac{g(t)-N}{t}
\end{equation}
which, by integration, yields Eq. (\ref{eq:Xiexp}). The proof ends up as above by expanding the exponential.



\end{document}